\documentclass[useAMS,usenatbib]{mn2e}
\usepackage{times}
\usepackage{graphicx}
\usepackage{amssymb,amsmath}
\usepackage[section]{placeins}
\newcommand\ion[2]{#1$\;${\scshape{#2}}}

\title[A pulsation zoo in KIC\,10139564.]{A pulsation zoo in the hot subdwarf B star KIC\,10139564 observed by {\it Kepler} \thanks{Based also on observations made with the Nordic Optical Telescope, operated on the island of La Palma jointly by Denmark, Finland, Iceland, Norway, and Sweden, in the Spanish Observatorio del Roque de los Muchachos of the Instituto de Astrofisica de Canarias.}
}
\author[A. S. Baran et al.]
{A. S. Baran $^{1,2}$\thanks{E-mail:sfbaran@cyf-kr.edu.pl},
M. D. Reed$^{1}$,
D. Stello$^3$,
R.H. {\O}stensen$^4$,
J.H. Telting$^5$,
E. Pak\v stien\.e$^6$,
\newauthor
S. J. O'Toole$^7$,
R. Silvotti$^{8}$,
P. Degroote$^{4,9}$,
S. Bloemen$^{4,9}$,
H. Hu$^{9,10}$,
V. Van Grootel$^{9,11}$,
\newauthor
B.D. Clarke$^{12}$,
J. Van Cleve$^{12}$,
S.E. Thompson$^{12}$
S.D. Kawaler$^{9,13}$
\\
$^{1}$ Department of Physics, Astronomy, and Materials Science, Missouri State University, Springfield, MO 65897, USA\\
$^{2}$ Uniwersytet Pedagogiczny, Obserwatorium na Suhorze, ul. Podchor\c{a}\.zych 2, 30-084 Krak\'ow, Polska\\
$^{3}$ Sydney Institute for Astronomy, School of Physics, University of Sydney, NSW 2006, Australia\\
$^{4}$ Instituut voor Sterrenkunde, K.U.\,Leuven, Celestijnenlaan 200D, 3001 Leuven, Belgium\\
$^{5}$ Nordic Optical Telescope, Apartado 474, 38700 Santa Cruz de La Palma, Spain\\
$^{6}$ Institute of Theoretical Physics and Astronomy, Vilnius University, Gostauto 12, Vilnius LT-01108, Lithuania\\
$^{7}$ Australian Astronomical Observatory, PO Box 296, Epping, NSW 1710, Australia\\
$^{8}$ INAF-Osservatorio Astronomico di Torino, Strada dell'Osservatorio 20, 10025 Pino Torinese, Italy\\
$^{9}$ Kavli Institute for Theoretical Physics, University of California Santa Barbara, USA \\
$^{10}$ Institute of Astronomy, The Observatories, Madingley Road, Cambridge CB3 0HA\\
$^{11}$ Institut	d' Astrophysique et de G\'eophysique, Universit\'e de Li\`ege, 17 All\'ee du 6 Ao{\^u}t, B 4000 Li\`ege, Belgium\\
$^{12}$ SETI Institute/NASA Ames Research Center, Moffett Field, CA 94035, USA\\
$^{13}$ Iowa State University, Department of Physics and Astronomy, 12 Physics Hall, Ames, IA 50011, USA\\
}
\begin{document}

\date{}

\pagerange{\pageref{firstpage}--\pageref{lastpage}} \pubyear{2010}

\maketitle

\label{firstpage}

\begin{abstract}
We present our analyses of 15\,months of {\it Kepler} data on KIC\,10139564. We detected 57 periodicities with a variety of properties not previously observed all together in one pulsating subdwarf B star. Ten of the periodicities were found in the low-frequency region, and we associate them with nonradial g-modes. The other periodicities were found in the high-frequency region, which are likely p-modes. We discovered that most of the periodicities are components of multiplets with a common spacing. Assuming that multiplets are caused by rotation, we derive a rotation period of 25.6$\pm$1.8\,days. The multiplets also allow us to identify the pulsations to an unprecedented extent for this class of pulsator. We also detect $l\geq2$ multiplets, which are sensitive to the pulsation inclination and can constrain limb darkening via geometric cancellation factors. While most periodicities are stable, we detected several regions that show complex patterns. Detailed analyses showed these regions are complicated by several factors. Two are combination frequencies that originate in the superNyquist region and were found to be reflected below the Nyquist frequency. The Fourier peaks are clear in the superNyquist region, but the orbital motion of {\it Kepler} smears the Nyquist frequency in the barycentric reference frame and this effect is passed on to the subNyquist reflections. Others are likely multiplets but unstable in amplitudes and/or frequencies. The density of periodicities also make KIC\,10139564 challenging to explain using published models. This menagerie of properties should provide tight constraints on structural models, making this subdwarf B star the most promising for applying asteroseismology.

To support our photometric analysis we have obtained spectroscopic radial-velocity measurements of KIC\,10139564 using low-resolution spectra in the Balmer-line region. We did not find any radial-velocity variation. We used our high S/N average spectrum to improve the atmospheric parameters of the sdB star, deriving T$_{\rm eff}$\,=\,31,859\,K and $\log g$\,=\,5.673\,dex.
\end{abstract}


\section{Introduction}
\label{introduction}
Hot subdwarf B stars (hereafter: sdB) are horizontal branch stars that consist of helium-rich cores surrounded by thin hydrogen envelopes. The helium core in each sdB star sustains nuclear burning of helium into carbon and oxygen, though in some cases the core may have exhausted helium at the center. Their progenitors are low-mass, main sequence stars, and each is likely to have undergone a core helium flash ($M \lesssim 2 M_{\odot}$). Once hydrogen in the core is exhausted, it evolves towards the tip of the red giant branch, where an sdB progenitor retains less than 0.01$M_{\odot}$ of hydrogen during (or prior to) the helium flash. Those red giant stars lie on the high $T_{\rm eff}$ ($\sim$30\,000K) end of the horizontal branch. This small fraction of hydrogen has an influence on the future evolution of sdB stars. Since hydrogen-shell burning cannot be sustained by a thin hydrogen envelope, the stars will move directly to the white dwarf cooling track instead of going through the asymptotic giant branch phase. Several mechanisms for producing this small residual hydrogen envelope involving single-star or binary evolution have been proposed;  {\it e.g.} \cite{dcruz96}; \cite{han02,han03}. The sdB stars have effective temperatures and surface gravities near 30\,000\,K and $\log g \sim$\,5.5, respectively.

It has been more than 14 years since the discovery of the first pulsating sdB star \citep[hereafter sdBV,][]{kilkenny97}. This star and several other stars discovered later were dominated by short period oscillations, which were identified as pressure (p-modes) modes by \cite{charp97}. They are driven in the envelopes and have allowed us to use asteroseismology to study the interiors of sdB stars. Several years later, \cite{green03} reported the discovery of another kind of variation in sdB stars. These brightness changes were smaller in amplitude and longer in period. According to the work done by \cite{fontaine03}, the new, longer period variations were assigned to gravity modes (g-modes) which dominate cooler, lower-gravity sdB stars. 

The g-modes in sdB stars probe deeper than p-modes and should allow us to study different parts of sdB stars.  Recently, \cite{reed11} used g-mode period spacings for mode identifications in sdB stars. It was found that $l$\,=\,1 period spacings were roughly 250\,sec while $l$\,=\,2 spacings were approximately 150\,sec. There are sdB stars pulsating in both p- and g-modes at the same time. These stars are the most favorable, since they allow us to probe larger parts of those stars. At first, only p-mode dominated, hybrid sdBV stars were detected from the ground, but the {\it Kepler} spacecraft has discovered several hybrid stars where the g-modes dominate across the entire amplitude spectrum \citep{reed10, baran11a}.

Nonradial pulsations can be described using spherical harmonics with quantized indices $n$, $l$, and $m$, where $n$ corresponds to the radial order of the mode, $l$ is the number of nodal lines on the surface, and $m$ is the number of those lines passing through the poles. Following the formalism given in \cite{ledoux51}, 
\begin{equation}
\sigma_{n,l,m}=\sigma_{n,l,0}+\Delta \sigma_{n,l,m}=\sigma_{n,l,0}+m\Omega(1-C_{n,l})
\label{ledoux}
\end{equation}
where $\sigma_{n,l,m}$ is the shifted frequency and $\sigma_{n,l,0}$ is the (degenerate) frequency without rotation. If the Ledoux constant $C_{n,l}$ is estimated and the multiplet splittings $\Delta \sigma_{n,l,m}$ are measured, we can derive the rotation frequency $\Omega$. It should be remembered that this equation is valid for slow, solid-body rotators with pulsation periods much shorter than the rotation period. In the absence of rotation, all modes of the same degree but with differing $m$ values are present in pulsating stars, but they are degenerate in frequency, though they differ in angular eigenfunctions. In this case, we only see the $\sigma_{n,l,0}$ frequency. However, rotation lifts the degeneracy in $m$, resulting in $(2l+1)$ periodicities with different frequencies. Periodicities with positive and negative $m$ are shifted in the opposite direction of $\sigma_{n,l,0}$. The $m$\,=\,0 frequency remains unaffected by rotation.

It is generally accepted that p-modes detected in sdB stars are of low order and of low degree. The Ledoux constant for such modes is typically close to zero. However, for the fundamental and first overtone $l$\,=\,2,\,3 modes, $C_{n,l}$ may have values up to 0.3 \citep{charp02}. This increase in the Ledoux constant creates a decrease in the frequency splittings. In the case of high-order g-modes, the Ledoux constant depends on the $l$ parameter according to the following equation: C$_{n,l}\lesssim(l^2+l)^{-1}$. It is around 0.5 for $l$\,=\,1 and close to 0.16 for $l$\,=\,2. If we assume a star rotates as a solid body, the ratio of the Ledoux constants between the p- and g-mode regions can be used to constrain the degree of the g-modes. A p/g splitting ratio of 0.55 indicates $l$\,=\,1 g-modes and a ratio of 0.81 indicates $l$\,=\,2 g-modes.

Observations of long period sdBV stars, including detection of new pulsators, are not easy from the ground. Relatively long period photometric variations (with only a few cycles per night) can be affected by variable sky transparency. To make a definitive determination of the pulsation frequencies requires extended photometric campaigns, preferably at several sites widely spaced in longitude to reduce day/night aliasing. However, {\it Kepler} provides long-duration, continuous, homogenous, evenly spaced time-series photometry, making it an ideal instrument for asteroseismology. The {\it Kepler} science goals, mission design, and overall performance are reviewed by \cite{borucki10}, \cite{koch10} and \cite{jenkins10}. This is currently the state-of-the-art way to observe variable stars. With this instrument, we can reach down to 19\,mag and still achieve better signal-to-noise (S/N) observations than from the ground.

\begin{table}
\begin{minipage}{83mm}
\centering
\caption{Details on KIC\,10139564. The last row reports the number of dates used in our analysis. Starting and ending points are given in BJD and are shifted to 2455000.0 epoch.}
\label{star}
\begin{tabular}{@{}cccc@{}}
\hline \hline
\multicolumn{2}{c}{R.A. [2000.0]} &  \multicolumn{2}{c}{DEC. [2000.0]}\\
\multicolumn{2}{c}{19 24 58.16} &  \multicolumn{2}{c}{+47 07 53.6}\\
\hline
$K_{\rm p}$ & T$_{\rm eff} [K]$ & $\log g$ [dex]  & $\log$ N$_{He}$/N$_{H}$ \\
16.1               & 31,859(126)       &  5.673(26)            & -2.201(36)           \\
\hline
start               & end                       &  length\,[days] & points  \\
276.485140 & 738.935300        & $\sim$462.5   & 625974 \\
\hline \hline
\end{tabular}
\end{minipage}
\end{table}

In this paper, we present analyses on KIC\,10139564 (2MASS\,19245816+4707536), a p-mode dominated sdBV star detected in {\it Kepler} data. Our analyses are completed on combined Q5 through Q9 data spanning 15\,months of continuous observations. Such long, continuous coverage -- never achieved before for compact pulsators -- significantly improves the S/N ratio in the amplitude spectra. This coverage allowed the detection of more frequencies and helped to improve mode identification, as compared to the preliminary results from the one-month survey data \citep{kawaler10}. In addition to {\it Kepler} data, we collected spectroscopic data, which we present in the next section.

\section{Spectroscopic data}
\label{spectroscopy}
On the nights of 31 May (2 spectra), 19 June (2) and 23 July 2010 (6), we obtained 10 spectra in total of KIC\,10139564 to monitor possible radial-velocity variations and, thus, check its binary status.  We used ALFOSC at the Nordic Optical Telescope with grism \#16 that yield spectra of the H$_\beta$--H$_{\theta}$ region with dispersion of 0.77\AA/pix and a resolution of 2.2\AA, with a median S/N\,=\,48 per pixel. The exposure times were 900 seconds. The spectra of 23 July were not taken consecutively so cover 0.3\,days.

The data were homogeneously reduced and analyzed. Standard reduction steps within IRAF were bias subtraction, removal of pixel-to-pixel sensitivity variations, optimal spectral extraction, and wavelength calibration based on arc lamp spectra. The target spectra and mid-exposure times were shifted to the barycentric frame of the solar system. The spectra were normalized to place the continuum at unity by comparing with a model spectrum for a star with similar physical parameters as we find for KIC\,10139564.

Radial velocities were derived with the FXCOR package in IRAF, with resulting errors of about 10\,km/s per spectrum. We used the lines of H$_\beta$, H$_\gamma$, H$_\delta$ and H$_\zeta$ to determine the radial velocities.

We find an average radial velocity of 1.0\,$\pm$2.9\,km/s and find all data points except the two from 31 May consistent within 1$\sigma$ with average radial velocity.  The two measurements of 31 May combine to a radial velocity of 15.5\,$\pm$7.1\,km/s.  We conclude that there is no clear evidence for orbital radial-velocity variations for periods shorter than half a day.

\begin{figure}
\includegraphics[width=83mm]{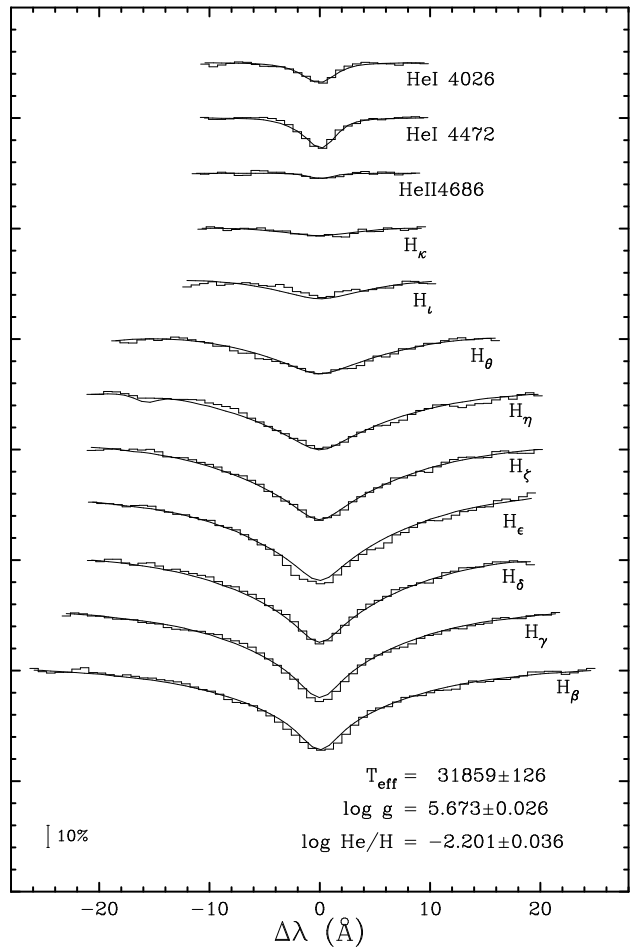}
\caption{The fit of an LTE model spectrum to a mean of 10 spectra obtained with the NOT/ALFOSC. Note that the H$_\epsilon$ line was kept out of the fit due to interstellar \ion{Ca}{ii} contamination.}
\label{spectrum}
\end{figure}

The 10 individual spectra were combined into an average spectrum with S/N$\sim$150, which was used for an improved determination of the atmospheric parameters of KIC\,10139564. We fitted the mean spectrum using the same method and grid of models as that used in \citet{osten10a}. \citep[For a description of these models, see][]{heber00}. The fit is shown in Figure\,\ref{spectrum}. The physical parameters resulting from the fitting procedure are listed in Table\,\ref{star}, and places KIC\,10139564 at a temperature some 600\,K lower and at a surface gravity 0.14\,dex lower than that obtained by \citet{osten10a} whose data were based on a single low S/N spectrum.

\section{Photometric data}
All data included in this paper were obtained through the {\it Kepler} Asteroseismic Science Consortium (KASC)\footnote{http://astro.phys.au.dk/KASC}. Since pixel data for Q6 and Q8 are still not available to us, we used flux measurements already calculated using default apertures defined by the in-house extraction pipeline \citep{jenkins10}. We have chosen to use the raw fluxes, which are calibrated for instrumental effects (bias, flat-field, cosmic rays) but not corrected for on-board systematics or flux excesses (often called contamination). 
Contamination does not influence our main goal, since we rely on periods and not amplitudes. We have corrected the data for on-board systematics to remove discontinuities in the data.

Table\,\ref{star} contains an observational report and the number of points used in our analysis. It is worth noting the unprecedented length of continuous coverage which cannot be obtained from the ground. The duty cycle is 92\%. We decided not to plot the photometric data, since they do not differ from those presented in the previous paper by \citet{kawaler10} other than being substantially longer in time.

We analyzed short cadence (SC) data. Each data point is a sum of nine integrations of 6.02\,s followed by a 0.52\,s readout, resulting in an SC sampling of 58.85\,s. This translates into a Nyquist frequency of 8496.356\,$\mu$Hz in the spacecraft reference frame. Time stamps given in that frame were corrected to the barycenter of the Solar System. This correction has smeared the Nyquist frequency because data are no longer evenly sampled.

In preparing data for Fourier analysis, we first removed all SC points flagged with errant times or fluxes. Then, all remaining points were subject to detrending which was done via a cubic spline fit calculated on 0.5\,day subsets and subtracted from the original data. Detrending with this interval will strongly suppress all frequencies below 12\,$\mu$Hz. Then we removed all outliers using 3.5$\sigma$ clipping before the data were subjected to Fourier analysis. We did not include the 30\,days of data from the survey phase, since they were too temporally separated from the data presented here, which could cause a complex pattern of peaks in the amplitude spectra and hinder our prewhitening process.

During Q5, Q6, Q7 and Q9, there were no safe-mode events, and cadences were only lost while the spacecraft was not in fine pointing (which occurred during monthly data transmissions to Earth). This makes these four quarters failure free. Unfortunately during Q8, the spacecraft experienced three safe-mode events: one which delayed the start of Q8 by about two weeks, one which caused a three-day gap, and one which prematurely ended Q8 observations 6 days early.

\section{Amplitude spectrum}
\label{amplspectra}
We used Fourier analysis to identify candidate pulsation frequencies.  Then we completed a nonlinear least-squares fit, including each periodicity in the form $A_{\rm i} \sin (\omega_{\rm i} t + \phi_{\rm i})$.  This iterative process followed the standard prewhitening procedure and continued until all peaks with amplitudes above a threshold of S/N\,=\,4 had been removed. While a threshold of S/N=4 is the commonly-used limit for sdB pulsations, it is not a strict limit and so frequencies near this limit (S/N$\leq$4.5) may be considered tentative, and are marked as such in Table\,\ref{10139564list}. Continued observations by Kepler may be used to confirm or refute them. The frequency resolution is 0.0375\,$\mu$Hz, defined as $1.5/{\rm T}$, where T is the time baseline of the data. We have coverage fifteen times longer than the 30\,days of survey data, and the noise level in the amplitude spectra are lower roughly by a factor of $\sqrt 15$\,=\,3.87.

\begin{figure*}
\includegraphics[width=180mm]{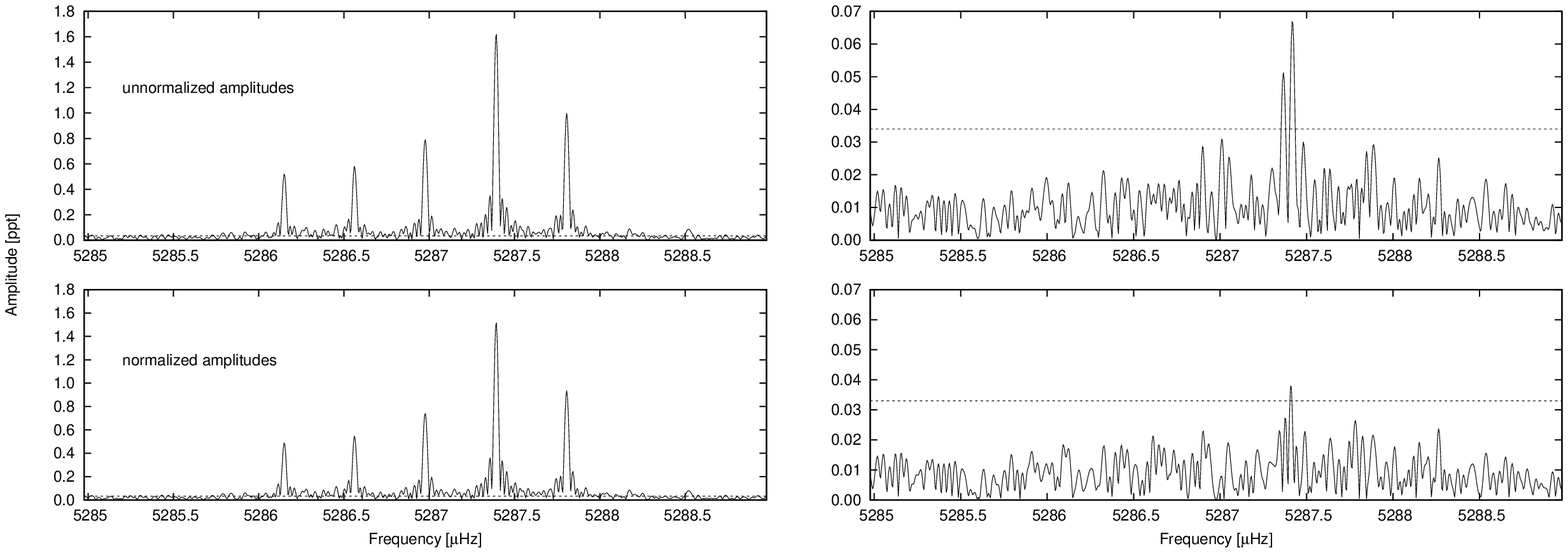}
\caption{Top panels show the unnormalized data and residuals while the bottom panels show the same for the amplitude-normalized data. The dashed horizontal line (here and in all other figures) represents the detection threshold of 0.033\,ppt for normalized and 0.034\,ppt for unnormalized data.}
\label{10139564f456a}
\end{figure*}

When the spacecraft is rolled every 3 months, the stars move onto different CCDs. Moreover, any deviation in pointing accuracy and other on-board systematics may cause different flux levels between quarters, which in turn leads to a different amplitude of light variation. Although we detrended the data using 0.5\,day subsets and 4$\sigma$ clipping, it is possible that instrumental systematics remain in the data. Such instrumental flux variations would decrease the effectiveness of prewhitening and manifest themselves as increased residuals and residual peaks with fork-like structures (multiple, closely-spaced peaks). Since many of the residuals in KIC\,10139564 do show fork-like structures we examined other options to further correct the data for instrumental flux changes. The best possible method would be to use a source external to the spacecraft, such as the star itself, for systematic flux corrections. Pulsation amplitudes can be used for this purpose, so long as the amplitudes themselves are stable with time. While many sdB pulsation amplitudes are variable, some do appear to be remarkably stable \citep[see \S 4.2 of][]{reed12}.
As such, we sought to correct these data using the amplitudes of the pulsations themselves. While this re-normalization could affect pulsation amplitudes, we are not interested in absolute amplitudes but rather the best possible frequency solution, one measure of which are residuals after prewhitening. The procedure we used was to split the data into month-long segments and fit the highest-amplitude pulsations. We then used those amplitudes to correct the original data on a month-by-month basis and re-prewhitened and analyzed the resultant residuals. We did this normalization one high amplitude pulsation frequency at a time. Only for the peak f$_{\rm 18}$ (Table\,\ref{10139564list}) did we achieve significant improvement. An example region is shown in Figure\,\ref{10139564f456a}. In this figure, the top panels show the unnormalized data and residuals while the bottom panels show the same for the amplitude-normalized data. The frequency solution is the same in both cases but the normalized residuals are smaller with less fork-like structure. The detection thresholds (horizontal dotted lines) for un- and normalized data are equal to 0.034 and 0.033\,ppt, respectively. Since the residuals were reduced for all frequencies, any amplitude changes in f$_{\rm 18}$ were most likely systematic and not intrinsic to that peak. For KIC10139564 this procedure did not change the frequency solution and so the sole benefit was in reducing several of the residual peaks below the detection threshold and removing some of the fork-like structures, which is a purely cosmetic result.

For this amplitude-normalized dataset we calculated the detection threshold as four times the average level of the residual spectrum in the range of 230\,$\mu$Hz to the Nyquist frequency. The detection threshold is 0.033\,ppt (parts per thousand). but can deviate locally by 3\% when using 500\,$\mu$Hz bins in signal-free regions. Please note that observed amplitudes are smeared by a finite sampling, leading to a decrease of 15-20\% for most periods and up to 35\% for the shortest periods.


KIC\,10139564 is the only short period dominated (p-mode) pulsator among sdBV stars detected by {\it Kepler}.  Based on 30\,days of survey data, \cite{kawaler10} noted that the amplitude spectrum of this star is dominated by several peaks between 160 and 200\,s (6200 and 5050\,$\mu$Hz) with the highest amplitude being about 9.9\,ppt. One single peak in the low-frequency region was also mentioned, but it had yet to be confirmed. Our analysis confirms all signals reported in the previous analysis. However, our frequency resolution is fifteen times better than in the survey data, allowing us to better resolve crowded peaks that lead to small changes in frequencies. In total, we have identified and successfully fit 57 periodicities with amplitudes above our detection threshold of 0.033\,ppt (Table\,\ref{10139564list}). Note that some signals prewhitened in the previous analysis were not prewhitened in ours. With our improved resolution, we found that the signal in some regions is very complex and difficult to fit. We discuss these regions in \S\ref{exotic}.

\begin{figure}
\includegraphics[width=83mm]{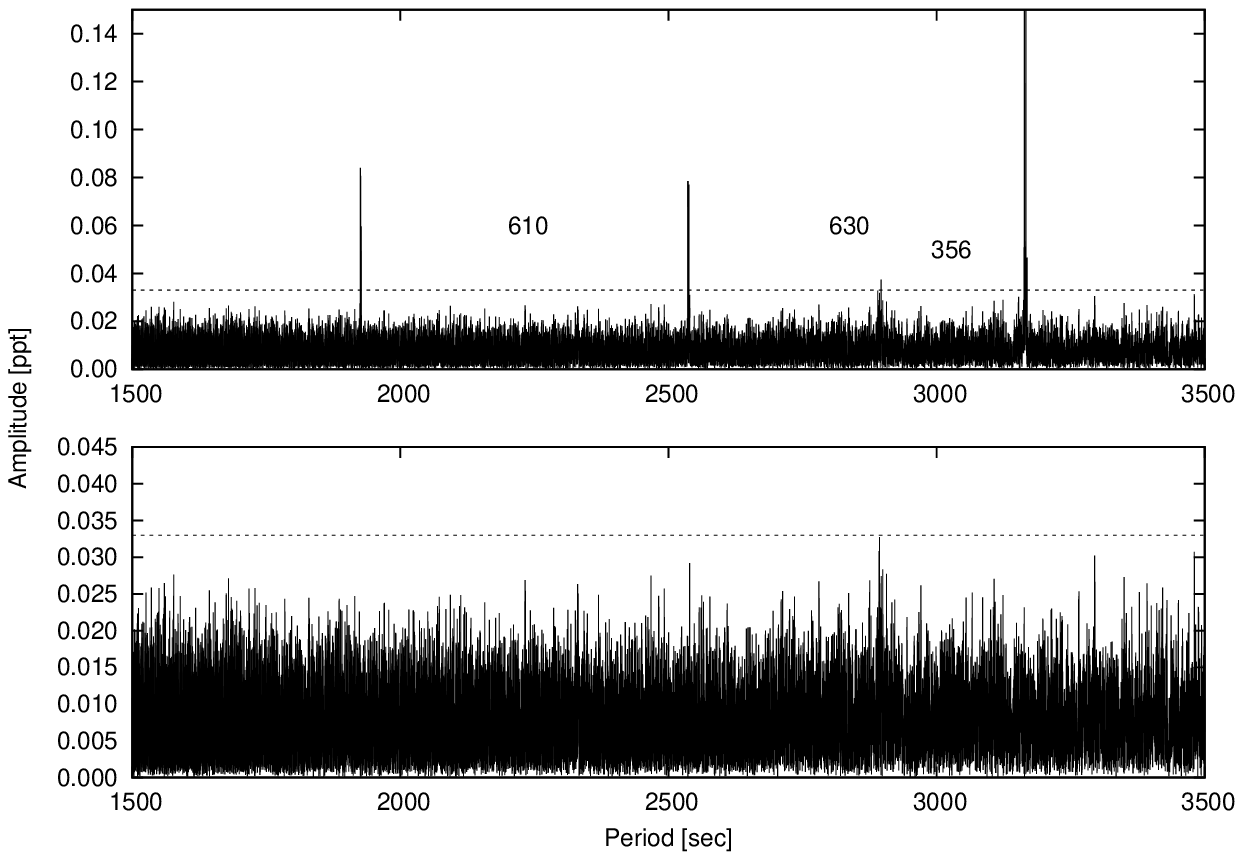}
\caption{Top panel: amplitude spectrum in the low-frequency region (note that abscissa axes are expressed in period instead of frequency and the top panel is truncated in amplitude). Three groups of peaks are evident. Numbers in the plots (here and in all other figures) denote separations between peaks, and they are given in abscissa axis units. 
Bottom panel: residual spectrum after removal of nine peaks (f$_{\rm A}$ to f$_{\rm J}$) in this region.}
\label{10139564lowf}
\end{figure}

\begin{table*}
\centering
\caption{The frequency list as a result of our prewhitening process for KIC\,10139564. The detection threshold is 0.033\,ppt. The numbers in parentheses are the 1-$\sigma$ fitting errors of the last digits. This list contains only peaks prewhitened from our data, which does not include some complex signal patterns described in \S\ref{exotic} and LC artifacts. Observed amplitudes are smeared by a finite sampling, leading to a decrease of 15-20\% in most periods and 35\% for the shortest periods. The asterisks denote tentative peaks with S/N$\leq$4.5. The single horizontal line separates low and high-frequency regions. Columns six and seven provide $l$ and {\it m} identification as a result of our analysis. This identification is based on full or partial multiplets. The last column, denoted K10, provides the notation of frequencies used in \citet{kawaler10}.}
\label{10139564list}
\begin{tabular}{@{}cccccccc@{}}
\hline \hline
ID   & Frequency [$\mu$Hz]  & Period[s]       & Amplitude [ppt] & S/N & $l$ & {\it m} & K10\\
\hline\hline
f$_{\rm A}$ & 315.5766(21) &   3168.803(21) &   0.050(8) &   6.0 & &-1 & \\
f$_{\rm B}$ & 315.8224(8) &   3166.336(7) &   0.138(8) &  16.6 & 1&0 & \\
f$_{\rm C}$ & 316.06559(31) &   3163.9002(31) &   0.339(8) &  40.8 & &+1 & f0\\
\cline{6-7}
f$_{\rm D}^*$ & 345.2455(28) &   2896.489(24) &   0.038(8) &   4.5 & 2&0 & \\
f$_{\rm E}^*$ & 345.9747(31) &   2890.385(26) &   0.034(8) &   4.1 & &+2 & \\
\cline{6-7}
f$_{\rm F}$ & 394.0282(14) &   2537.889(9) &   0.078(8) &   9.4 & 1&0 & \\
f$_{\rm G}$ & 394.2876(14) &   2536.220(9) &   0.079(8) &   9.5 & &+1 & \\
\cline{6-7}
f$_{\rm H}$ & 518.8984(19) &   1927.160(7) &   0.057(8) &   6.8 & &-1 & \\
f$_{\rm I}$ & 519.1540(14) &   1926.2106(50) &   0.079(8) &   9.4 & 1&0 & \\
f$_{\rm J}$ & 519.4022(13) &   1925.2904(49) &   0.081(8) &   9.7 & &+1 & \\
\hline
f$_{\rm 1}^*$ &2212.6076(29) &    451.95543(6) &   0.037(8) &   4.4 & & & \\
f$_{\rm 2}^*$ &3276.9340(30) &    305.163302(28) &   0.035(8) &   4.2 & & & \\
\cline{6-7}
f$_{\rm 3}$ &3540.4565(22) &    282.44945(18) &   0.048(8) &   5.8 & 2&0 & \\
f$_{\rm 4}$ &3541.4320(19) &    282.37165(15) &   0.056(8) &   6.8 & &+2 & \\
\cline{6-7}
f$_{\rm 5}$ &5048.7427(21) &    198.06912(8) &   0.052(8) &   6.2 & &-2 & \\
f$_{\rm 6}$ &5049.7109(23) &    198.03114(9) &   0.046(8) &   5.5 & 2&0 & \\
\cline{6-7}
f$_{\rm 7}^*$ &5052.6025(31) &    197.91781(12) &   0.034(8) &   4.1 & & & \\
\cline{6-7}
f$_{\rm 8}$ &5286.14933(22) &    189.173619(8) &   0.478(8) &  57.4 & &-2 &  f1\\
f$_{\rm 9}$ &5286.56176(20) &    189.158861(7) &   0.544(8) &  65.4 & &-1 & \\
f$_{\rm 10}$ &5286.97660(15) &    189.144018(5) &   0.723(8) &  86.9 & 2&0 & \\
f$_{\rm 11}$ &5287.39188(7) &    189.1291628(25) &   1.521(8) & 182.9 & &+1 &  f2\\
f$_{\rm 12}$ &5287.80529(12) &    189.1143764(42) &   0.899(8) & 108.1 & &+2 & f3\\
\cline{6-7}
f$_{\rm 13}$ &5410.6980(9) &    184.819039(30) &   0.121(8) &  14.6 & &-2 & \\
f$_{\rm 14}$ &5411.60111(50) &    184.788195(17) &   0.210(8) &  25.3 & 2&0 & \\
f$_{\rm 15}$ &5412.5138(7) &    184.757036(25) &   0.143(8) &  17.2 & &+2 & \\
\cline{6-7}
f$_{\rm 16}$ &5413.39487(29) &    184.726964(10) &   0.363(8) &  43.6 & 1&0 & f4\\
f$_{\rm 17}$ &5413.8205(9) &    184.712442(29) &   0.124(8) &  14.9 & &+1 & \\
\cline{6-7}
f$_{\rm 18}$ &5472.86225(3) &    182.7197459(10) &   3.712(8) & 446.4 & 0&0 & f7\\
\cline{6-7}
f$_{\rm 19}$ &5570.0278(10) &    179.532318(31) &   0.111(8) &  13.4 & &-3 & \\
f$_{\rm 20}^*$ &5570.4811(31) &    179.51771(10) &   0.034(8) &   4.1 & &-2 & \\
f$_{\rm 21}$ &5571.3956(15) &    179.488243(49) &   0.070(8) &   8.4 & 3&0 & \\
f$_{\rm 22}$ &5572.2929(23) &    179.45934(8) &   0.045(8) &   5.5 & &+2 & \\
f$_{\rm 23}^*$ &5572.7357(29) &    179.44508(9) &   0.037(8) &   4.5 & &+3 & \\
\cline{6-7}
f$_{\rm 24}$ &5616.4852(26) &    178.04730(8) &   0.040(8) &   4.8 & & & \\
f$_{\rm 25}^*$ &5618.9672(29) &    177.96865(9) &   0.036(8) &   4.3 & & & \\
\cline{6-7}
f$_{\rm 26}$ &5708.5701(21) &    175.17522(6) &   0.051(8) &   6.2 & & & \\
f$_{\rm 27}$ &5709.3992(28) &    175.14978(8) &   0.039(8) &   4.6 & & & \\
\cline{6-7}
f$_{\rm 28}$ &5746.6089(13) &    174.015672(41) &   0.079(8) &   9.5 & &-3 & \\
f$_{\rm 29}$ &5746.9531(26) &    174.00525(8) &   0.041(8) &   4.9 & &-- & \\
f$_{\rm 30}$ &5747.0936(6) &    174.000993(17) &   0.186(8) &  22.3 & &-2 & \\
f$_{\rm 31}$ &5748.0723(8) &    173.971367(23) &   0.142(8) &  17.1 & 3&0 & \\
f$_{\rm 32}$ &5749.0647(12) &    173.941338(35) &   0.091(8) &  10.9 & &+2 & \\
\cline{6-7}
f$_{\rm 33}$ &5760.16218(1) &    173.6062229(4) &   7.950(8) & 955.8 & &-1 & f8\\
f$_{\rm 34}$ &5760.58982(4) &    173.5933352(12) &   2.741(8) & 329.6 & 1&0 & \\
f$_{\rm 35}$ &5761.00886(2) &    173.5807085(5) &   5.968(8) & 717.6 & &+1 & f9\\
\cline{6-7}
f$_{\rm 36}$ &5839.3800(23) &    171.25106(7) &   0.047(8) &   5.6 &4 &-2 & \\
f$_{\rm 37}$ &5841.5754(20) &    171.18670(6) &   0.054(8) &   6.5 & &+4 & \\
\cline{6-7}
f$_{\rm 38}$ &6001.1015(6) &    166.636074(18) &   0.166(8) &  20.0 & &-1 & \\
f$_{\rm 39}$ &6001.4733(7) &    166.625753(19) &   0.154(8) &  18.5 & 1&0 & f12\\
\cline{6-7}
f$_{\rm 40}$ &6076.2276(8) &    164.5757965(23) &   0.126(8) &  15.1 & & & f14\\
f$_{\rm 41}$ &6106.2252(20) &    163.767297(50) &   0.053(8) &   6.4 & & & \\
\cline{6-7}
f$_{\rm 42}$ &6234.71348(13) &    160.3922943(32) &   0.850(8) & 102.3 & 1&0 & f16\\
f$_{\rm 43}$ &6235.03682(33) &    160.383977(9) &   0.323(8) &  38.8 & &+1 & \\
\cline{6-7}
f$_{\rm 44}$ &7633.6773(13) &    130.998464(23) &   0.081(8) &   9.7 & & & f19\\
\cline{6-7}
f$_{\rm 45}$ &8117.2934(22) &    123.193773(33) &   0.048(8) &   5.8 & &-2 & \\
f$_{\rm 46}$ &8118.7808(16) &    123.171203(25) &   0.065(8) &   7.8 & 2&+1 & \\
f$_{\rm 47}$ &8119.2439(20) &    123.164179(30) &   0.054(8) &   6.5 & &+2 & \\
\hline \hline
\end{tabular}
\end{table*}

\section{Discussion}

\subsection{Signal in the low-frequency region}
In the low-frequency region, we have detected ten peaks (f$_{\rm A}$ to f$_{\rm J}$). All low-frequency peaks are listed in the upper part of Table\,\ref{10139564list} and are shown in Figure\,\ref{10139564lowf}. We associated all of them with g-modes.

\begin{figure}
\includegraphics[width=83mm,height=210mm]{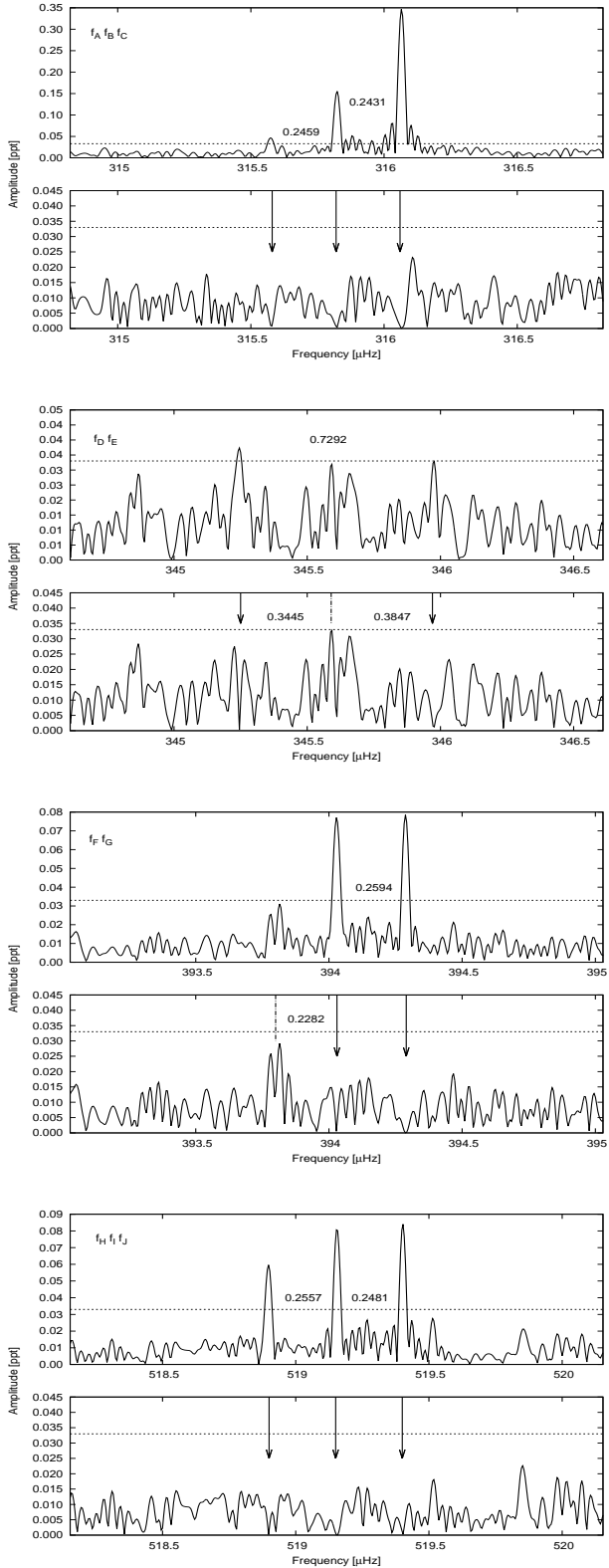}
\caption{Zoomed amplitude spectra for the g-mode region. Each pair of panels contains the original (top) and residual spectra (bottom). Prewhitenned peaks are marked with arrows. Frequency separations, which include two members with S/N$<$4 (marked with vertical dash-dotted lines), are labeled in the residual panels; those separations are not precise and may differ by up to 0.03\,$\mu$Hz.}
\label{10139564f27}
\end{figure}

\begin{figure}
\includegraphics[width=83mm]{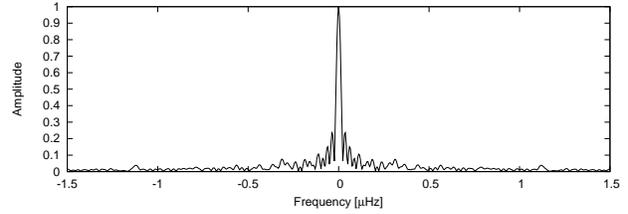}
\caption{The window function calculated from Q5-Q9 data.}
\label{window}
\end{figure}

In each region, the signal is split into two or three significant peaks as seen in Figure\,\ref{10139564f27}. The frequency separations between peaks are similar and includes periodicities with amplitudes that have S/N$<$4, shown in Figure\,\ref{10139564f27}. In Figure\,\ref{window}, we plot the window function to show that the splitting is not a result of the aliasing pattern. Since rotation can remove degeneracies in pulsation frequencies of modes with the same $l$ but different {\it m} value, we suggest that these are rotationally split multiplets. With two or three peaks observed, we can consider them to be $l$\,=\,1 triplets; however with more data, it is possible they may turn into quintuplets. Two frequencies f$_{\rm D}$ and f$_{\rm E}$ are different. They show larger splittings of about 0.72\,$\mu$Hz. An insignificant frequency (S/N=3.8), marked with a vertical dash-dotted line, between them could create a multiplet with a splitting of 0.36\,$\mu$Hz on average. This value is expected for a quintuplet since the Ledoux constant is 67\% smaller for $l$\,=\,2 modes. These frequencies also overlap an artifact listed in Data Release Notes 12 Notes as ``wide'' and it is possible they are not intrinsic to the star. However, that artifact was discovered by one of us (ASB) in Q0 data. In addition, our two peaks do not resemble the artifact found in Q0, so we infer them to be produced by the star.


All peaks except f$_{\rm D}$ and f$_{\rm E}$ are organized into three groups with the separations between them of 610 and 630\,sec, respectively. This is not surprising, since g-modes should be equally spaced in period in accordance with asymptotic theory. In KIC\,10139564 we could have a sequence of $l$\,=\,2 modes but only with every fourth multiple (620/4=155) detected. However we assigned them as $l$\,=\,1 modes based on their multiplet structures, and that makes them a sequence with spacings of (620/2=)310 or (620/3=)207\,sec, if only the second or third multiples were detected, respectively. While the $l$\,=\,1 period spacings would not be consistent with other pulsating sdB stars, those stars are g-mode dominated while KIC\,10139564 is p-mode dominated and the frequency splittings indicate these g-modes are likely $l$\,=\,1.

\begin{figure}
\includegraphics[width=83mm]{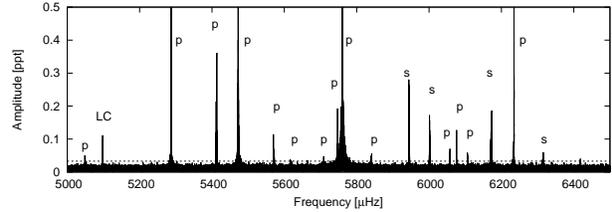}
\caption{The high-frequency region of KIC\,10139564. This plot is truncated at 0.5\,ppt to show low-amplitude peaks. Labels help to see how this region is organized. We found at least one strong LC artifact, many peaks we could prewhiten (p) and complex signal (s) which we left untouched.}
\label{10139564highf}
\end{figure}

\subsection{Signal in the high-frequency region}
\label{high}
The high-frequency region contains 47 significant peaks. We show this region in Figure\,\ref{10139564highf}. The highest peak has an amplitude of 7.950\,ppt and four other peaks are between one and six ppt. Most of the peaks, however, are of low amplitudes, barely exceeding S/N\,=\,10. Signals prewhitened from our data are denoted as {\it p}, LC means long cadence artifacts and {\it s} stands for complex signal regions, which we were not able to remove from the data using standard techniques. LC artifacts are peaks at the long cadence frequency (566.41\,$\mu$Hz) and its harmonics up to the Nyquist frequency. They were identified during the first year of the {\it Kepler} mission, and their frequencies are given in the {\it Kepler Instrument Handbook} ({\it KIH}). Besides LC artifacts, {\it Kepler} data may also contain other spurious peaks. Some of them have been identified in {\it KIH}, but there are others that are still not documented very well. \cite{baran11b} reported another set of artifact signals detected in many stars analyzing {\it Kepler} public data. It is also possible that artifacts may come and go, and a frequency of a specific artifact may also change. In the amplitude spectra we have detected all LC harmonics starting from the 6$^{\rm th}$ up to the 14$^{\rm th}$, which is the Nyquist frequency.

\begin{figure}
\includegraphics[width=83mm]{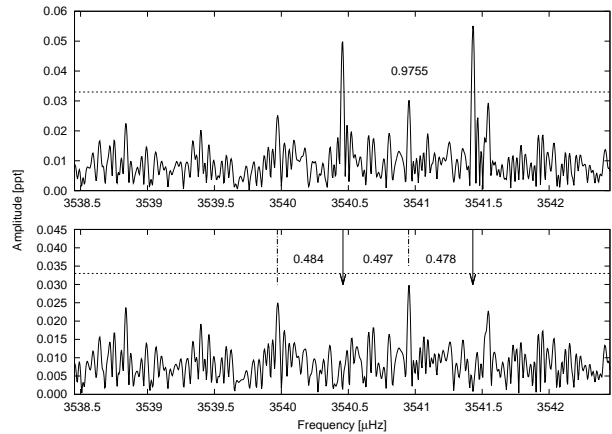}
\caption{The frequency region around 3540\,$\mu$Hz. We detected two significant peaks f$_{\rm 3}$ and f$_{\rm 4}$, but there are two other peaks slightly below the detection limit which fit into the evenly spaced structure.}
\label{10139564f305}
\end{figure}

\begin{figure}
\includegraphics[width=83mm]{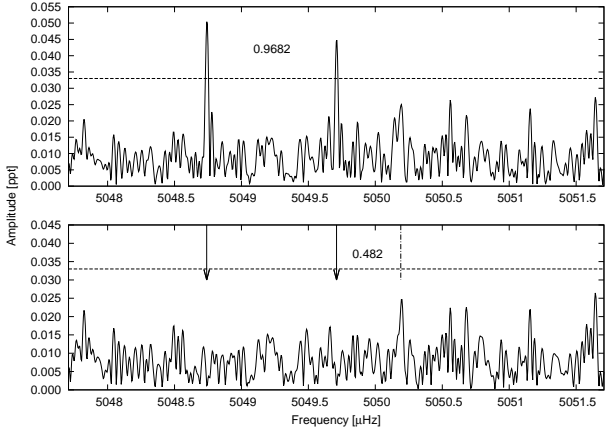}
\caption{The frequency region around 5049\,$\mu$Hz. Similar to Figure\,\ref{10139564f305}, we found two low-amplitude peaks f$_{\rm 5}$ and f$_{\rm 6}$ and another peak slightly below the significance limit. It is worth noting that separations between these peaks are comparable to the peaks in the Figure\,\ref{10139564f305}.}
\label{10139564f436}
\end{figure}

The first few periodicities (f$_{\rm 1}$ to f$_{\rm 4}$) are well separated from the bulk of p-modes. Frequencies f$_{\rm 1}$ and f$_{\rm 2}$ appeared to be singlets. At 3540\,$\mu$Hz, we fit two significant peaks (f$_{\rm 3}$ and f$_{\rm 4}$) as seen in Figure\,\ref{10139564f305}. In addition, two other peaks seem likely to represent real signal (particularly the peak close to 3541\,$\mu$Hz with S/N\,=\,3.6), but we need more data to confirm this. Both significant peaks have low amplitudes and the separation between them is 0.9755\,$\mu$Hz. Estimating frequencies for the two other peaks and calculating separations reveals that all four peaks create an almost equally spaced quadruplet. We found a similar case at 5049\,$\mu$Hz shown in Figure\,\ref{10139564f436}. We detected two low-amplitude, yet significant, peaks (f$_{\rm 5}$ and f$_{\rm 6}$) separated by 0.9682\,$\mu$Hz with another insignificant (S/N=3) peak 0.482\,$\mu$Hz away from f$_{\rm 6}$. The separations between peaks in the group at 3540 and 5049\,$\mu$Hz are similar. The peak at f$_{\rm 7}$ stands alone and barely exceeds the detection threshold. If f$_{\rm 7}$ is a part of a multiplet, other components are currently undetected.

\begin{figure}
\includegraphics[width=83mm]{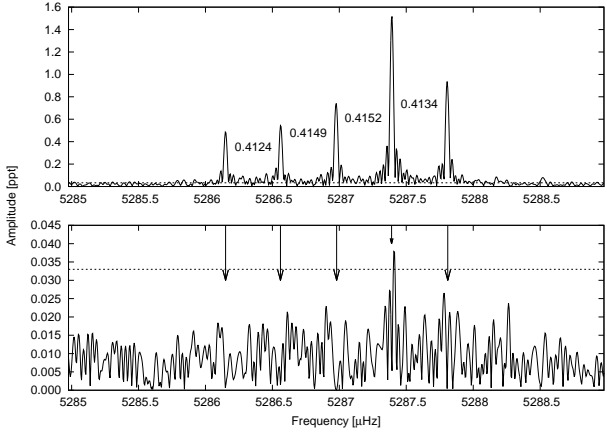}
\caption{The frequency region around 5287\,$\mu$Hz. This region contains a clearly detected quintuplet (f$_{\rm 8}$ to f$_{\rm 12}$).}
\label{10139564f456}
\end{figure}

\begin{figure}
\includegraphics[width=83mm]{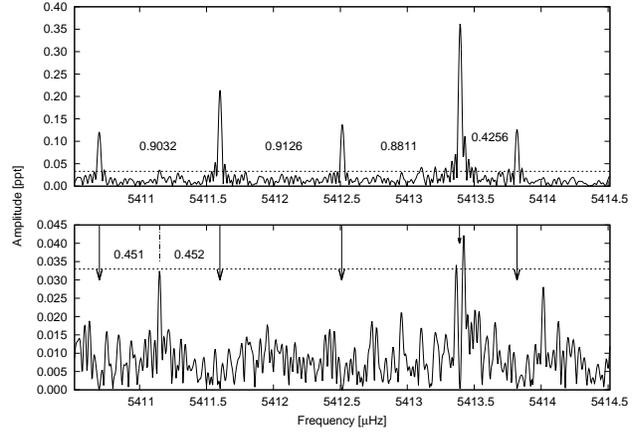}
\caption{The frequency region around 5412\,$\mu$Hz. This region shows five peaks (f$_{\rm 13}$ to f$_{\rm 17}$), but unlike the previous region, they do not create an evenly spaced obvious quintuplet.}
\label{10139564f467}
\end{figure}

The region presented in Figure\,\ref{10139564f456} contains a clearly resolved quintuplet (f$_{\rm 8}$ to f$_{\rm 12}$) with separations of $\sim$\,0.41\,$\mu$Hz, comparable to those in the previous p-mode regions. As may be seen in the residual spectrum (bottom panel), all peaks prewhitened with almost no residual signal left.

Figure\,\ref{10139564f467} shows the region near 5412\,$\mu$Hz that also contains five significant peaks (f$_{\rm 13}$ to f$_{\rm 17}$). However, these peaks do not create an evenly spaced structure, yet the separations look familiar. One separation is $\sim$\,0.4256\,$\mu$Hz, while the other three appear approximately twice that near 0.9\,$\mu$Hz. In addition, after all significant peaks were removed, we found another peak (with S/N=3.9) between f$_{\rm 13}$ and f$_{\rm 14}$ that may fit into this sequence. Two possible interpretations exist. Either all the peaks create one large $l>$3 multiplet with separations of $\sim$0.44\,$\mu$Hz or, as we prefer, f$_{\rm 13}$ to $f_{\rm 15}$ are part of a quintuplet, f$_{\rm 16}$ and f$_{\rm 17}$ are part of a triplet and the two multiplets just happen to be separated by $\sim$0.9\,$\mu$Hz.

\begin{figure}
\includegraphics[width=83mm]{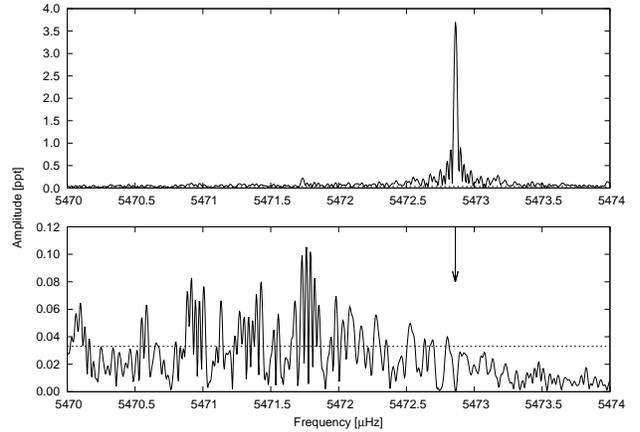}
\caption{The frequency region around 5473\,$\mu$Hz with only one peak f$_{\rm 18}$ and a wide, low amplitude signal at lower frequency.}
\label{10139564f472}
\end{figure}

\begin{figure}
\includegraphics[width=83mm]{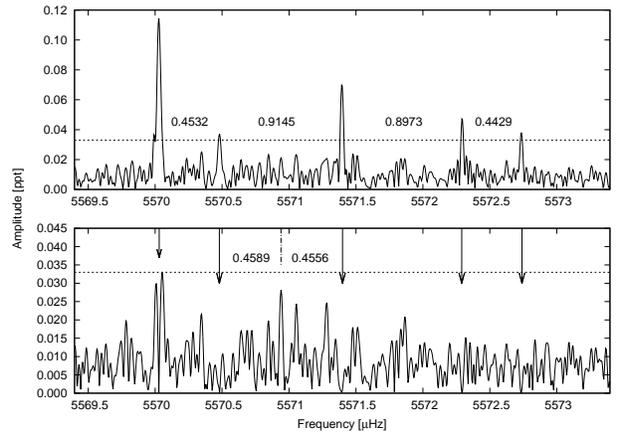}
\caption{The frequency region around 5572\,$\mu$Hz. We detected five peaks f$_{\rm 19}$ to f$_{\rm 23}$ with separations comparable to the previous regions. This may represent a septuplet.}
\label{10139564f481}
\end{figure}

The peak f$_{\rm 18}$ shown in Figure\,\ref{10139564f472} is different. It seems to be single with some low-amplitude complex signal at slightly lower frequencies. The amplitude of this peak was used to normalize data between individual months, which significantly removed residuals. This is the only single high amplitude peak we have detected in our data, and its lack of multiplets, despite its high amplitude, leads us to conclude it is a radial mode. In the residual spectrum (bottom panel of Figure\,\ref{10139564f472}), we notice an obvious and very wide signal excess. We did not try to prewhiten any peaks from this "forest". It resembles other regions of peaks between 5900 and 6400\,$\mu$Hz.

\begin{figure}
\includegraphics[width=83mm]{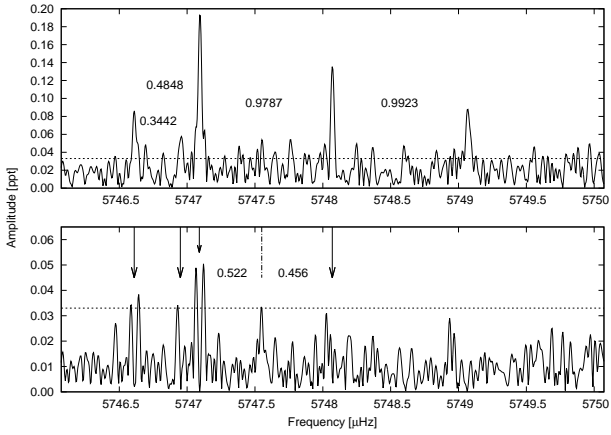}
\caption{The frequency region near 5748\,$\mu$Hz with five peaks (f$_{\rm 28}$ to f$_{\rm 32}$). Peaks f$_{\rm 30}$ to f$_{\rm 32}$ might be part of a quintuplet, since the separation between them is twice the smallest splitting in other multiplets.}
\label{10139564f496}
\end{figure}

The signal at 5572\,$\mu$Hz, shown in Figure\,\ref{10139564f481}, might be of special interest. We detect five peaks, f$_{\rm 19}$ to f$_{\rm 23}$, with an average spacing of 0.45\,$\mu$Hz assuming that two peaks just below the detection threshold are intrinsic to KIC\,10139564. The residual spectrum has a peak at the expected frequency to complete the sequence but it is not significant (S/N=3.41). This region could contain an $l$\,=\,3 septuplet. Thus far, the detection of l$>$2 modes was generally regarded as unlikely, since these modes suffer from large surface cancellation effects, making their observed amplitudes greatly reduced, assuming the intrinsic amplitudes to be similar. With {\it Kepler} data we can reach down to the ppm level, making the detection of septuplets possible. More data are definitely needed to confirm the nature of the peaks in this region, but it is a good candidate for the first detection of $l \geq$\,3 modes in any sdB star.

\begin{figure}
\includegraphics[width=83mm]{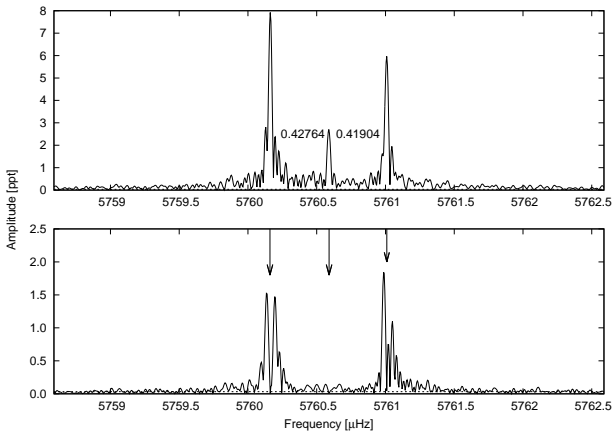}
\caption{The region around 5760\,$\mu$Hz which contains a triplet (f$_{\rm 33}$ to f$_{\rm 35}$) with splittings comparable to other regions. This would indicate that the highest amplitude peak is \emph{not} a radial mode.}
\label{10139564f497}
\end{figure}

We do not show the next region at 5618\,$\mu$Hz, since it only contains two low amplitude peaks f$_{\rm 24}$ and f$_{\rm 25}$ separated by 2.482\,$\mu$Hz. It may be the case of another septuplet with the two peaks being the first and last in a sequence of seven modes separated approximately by 0.42\,$\mu$Hz. The residual spectrum shows one peak near the detection threshold and fits the potential sequence well. Similarly, only two low-amplitude peaks f$_{\rm 26}$ and f$_{\rm 27}$ were found at 5709\,$\mu$Hz. More data are necessary to determine the nature of these two regions.

\begin{figure}
\includegraphics[width=83mm]{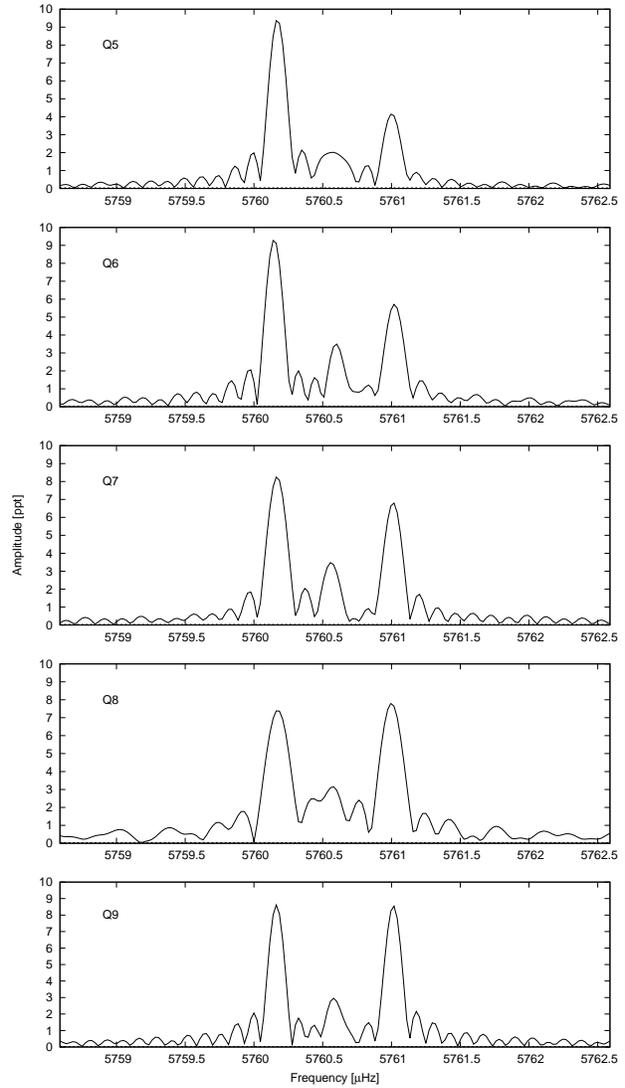}
\caption{The amplitude spectra of the highest amplitude periodicities (f$_{\rm 33}$ to f$_{\rm 35}$) calculated from data split into quarters. Notice the amplitude shift between the outer frequencies. In Q8 the peak on the right is slightly higher than the one on the left.}
\label{10139564f497ampl}
\end{figure}

Similar to the region shown in Figure\,\ref{10139564f481}, we have detected a triplet at 5748\,$\mu$Hz (f$_{\rm 30}$ to f$_{\rm 32}$) with separations of 0.98\,$\mu$Hz on average (Figure\,\ref{10139564f496}). This could easily fit a quintuplet with intrinsic splittings of 0.49\,$\mu$Hz, similar to other multiplets. In addition, f$_{\rm 28}$ and the peak (S/N$\sim$4) at 5747.55\,$\mu$Hz could potentially fit this sequence. Another option could be that f$_{\rm 28}$ to f$_{\rm 29}$ are part of a different multiplet with shorter separations of 0.3442\,$\mu$Hz which is similar to the multiplets found at 5840 and 6235\,$\mu$Hz.

The next p-mode region is shown in Figure\,\ref{10139564f497}. It is centered around 5760\,$\mu$Hz and contains three significant peaks (f$_{\rm 33}$ to f$_{\rm 35}$). Since radial modes suffer no geometric cancellation, it is usually anticipated that the highest amplitude modes are, therefore, radial. This does not seem to be the case for KIC\,10139564. All three peaks are spaced by the distance of roughly 0.42\,$\mu$Hz, forming a triplet. These spacings in the highest amplitude frequencies provide assurance that this frequency spacing, which is common to many regions of KIC\,10139564, is intrinsic and related to rotationally split multiplets. The residual spectrum shown in the bottom panel of Figure\,\ref{10139564f497} indicates that all those peaks are unstable in their amplitudes/phases. In Figure\,\ref{10139564f497ampl} we show how the peaks change their amplitudes between quarters; in Q8 the amplitude of the right peak was the highest.

\begin{figure}
\includegraphics[width=83mm]{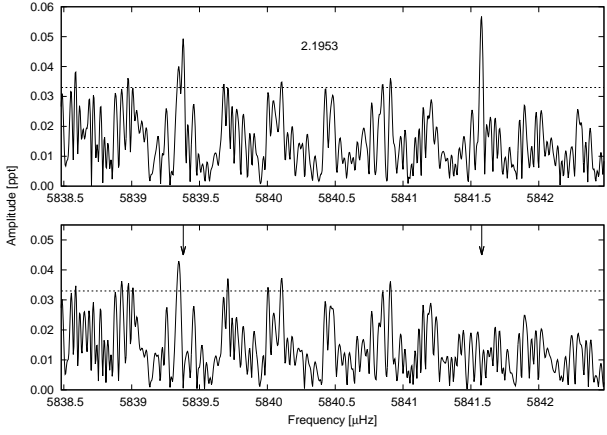}
\caption{The frequency region around 5840\,$\mu$Hz. We detected two peaks f$_{\rm 36}$ and f$_{\rm 37}$.}
\label{10139564f504}
\end{figure}

The region at 5840\,$\mu$Hz contains low amplitude peaks f$_{\rm 36}$ and f$_{\rm 37}$ with very complex shapes shown in Figure\,\ref{10139564f504}. They appear to be suffering from amplitude/phase variations or as unresolved frequencies, which makes our fitting process difficult and our estimation of their frequencies and amplitudes less precise. Regardless, we have fitted f$_{\rm 36}$ and f$_{\rm 37}$ with a separation of 2.1953\,$\mu$Hz. Other peaks in this region cannot be fitted. We will discuss this region in detail in \S\ref{exotic}.

\begin{figure}
\includegraphics[width=83mm]{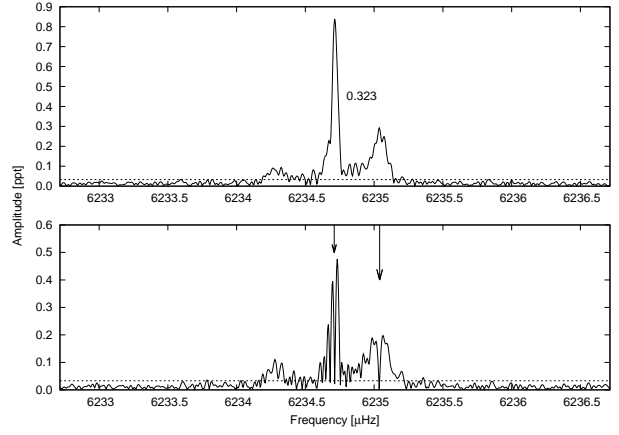}
\caption{The frequency region around 6235\,$\mu$Hz with two peaks detected (f$_{\rm 42}$ and f$_{\rm 43}$). The wide peaks possibly indicate amplitude/phase variation.}
\label{10139564f538}
\end{figure}

The region at 6001\,$\mu$Hz is one with complex signal patterns and is described in Sect.\ref{exotic}. Two frequencies f$_{\rm 38}$ and f$_{\rm 39}$ look relatively high and stable in amplitudes, so we fitted and prewhitened them (listed in Table\,\ref{10139564list}). The next two regions around 6076\,$\mu$Hz (f$_{\rm 40}$) and 6106\,$\mu$Hz (f$_{\rm 41}$) are not shown since we fitted just one significant low amplitude peak each. The region at 6235\,$\mu$Hz shows three significant peaks (Figure\,\ref{10139564f538}), but their structure is very complex. We could easily prewhiten two of them (f$_{\rm 42}$ and f$_{\rm 43}$), leaving the third one untouched. The residual spectrum clearly shows amplitude and/or phase variation of all peaks. The smallest (lowest frequency) signal has a fork-like appearance, which may mean it is unresolved in the original spectrum, and we decided not to make an attempt to remove it. The spacing between the two prewhitened peaks is slightly smaller than the common value of 0.45\,$\mu$Hz, as derived in other regions. This deviation can be explained by a difference in the Ledoux constants (\S\ref{introduction}). The remaining signal appears at exactly the right frequency to fit a triplet, leading us to conclude this region is another rotational multiplet. There is another significant signal in the amplitude spectrum around 7633\,$\mu$Hz with a single frequency f$_{\rm 44}$ detected.

\begin{figure}
\includegraphics[width=83mm]{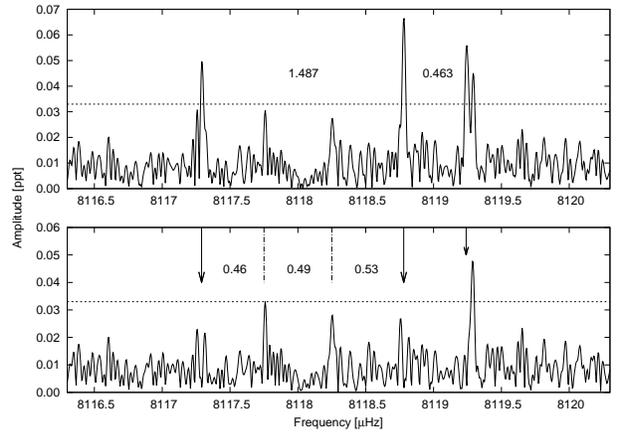}
\caption{The frequency region around 8118\,$\mu$Hz with three peaks detected (f$_{\rm 45}$ to f$_{\rm 47}$). The rightmost peak is very wide and is likely unresolved, as the residual seems to confirm. Including the two near the detection limit, this region form a quintuplet.}
\label{10139564f701}
\end{figure}

The region at 8118\,$\mu$Hz is the highest frequency region with three peaks f$_{\rm 45}$ to f$_{\rm 47}$ that we could easily prewhiten. In fact, it is very close to the Nyquist frequency, leaving some doubts whether this region is mirrored across that frequency. If so, the real signal could appear at approximately 8874\,$\mu$Hz instead. Similar to the first region at 3540\,$\mu$Hz, this region is also well separated from the bulk of signals we detected in the amplitude spectrum. We show this region in Figure\,\ref{10139564f701}. It seems obvious that five peaks are spaced by the common splitting (two just below the detection threshold), and we interpret them as another $l$\,=\,2 quintuplet.

\subsection{Multiplets}
\label{multiplets}
We expect all nonradial modes of a slowly rotating star to be split according to the Ledoux equation (Eq.\,\ref{ledoux}). This is indeed what we see in the amplitude spectrum of KIC\,10139564. Most peaks are split with roughly consistent splitting among modes of the same kind (p- or g-modes).

\begin{figure}
\includegraphics[width=83mm]{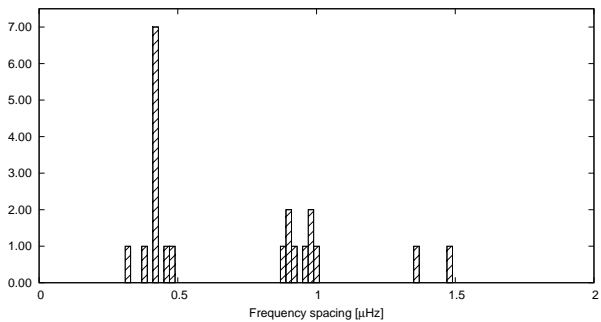}
\caption{A histogram of frequency spacings in the p-mode region.}
\label{spl_hist}
\end{figure}

In Figure\,\ref{spl_hist} we show spacings between p-modes with S/N$>$5.5. Most group at approximately 0.45\,$\mu$Hz, with a few having double that spacing at 0.9\,$\mu$Hz and two with three times that spacing near 1.4\,$\mu$Hz. We interpret the group with the smallest spacings to arise from modes of consecutive $m$ values ($\Delta m\,=\,1$), the next group due to $\Delta m\,=\,2$, and the final pair due to $\Delta m\,=\,3$. Using the average splitting between p-modes of 0.452$\pm$0.030\,$\mu$Hz) and $C_{n,l}$\,=\,0 (splittings below 0.4\,$\mu$Hz were not taken into account since we suspect they are caused by a non-zero Ledoux constant), we derived an average rotation period of 25.6$\pm$1.8\,days.

Multiplets at low frequencies have average splittings of 0.2504$\pm$0.0019 and 0.3646$\pm$0.0042\,$\mu$Hz. As mentioned in \S\ref{introduction} the ratio of the Ledoux constant between the p- and g-mode regions can be used to constrain the degree of the g-modes. The smaller splitting has a ratio of 0.55 and the larger splitting has a ratio of 0.81 which matches the Ledoux ratio if the g-modes are $l$\,=\,1 and $l$\,=\,2, respectively.
 
\cite{baran09} reported the discovery of a symmetrical but not evenly spaced quintuplet in the sdBV star Balloon\,090100001 (Bal09) and explained this uneven spacing as differential rotation on the surface. In KIC\,10139564, we detected the quintuplet f$_{\rm 8}$ to f$_{\rm 12}$, which is symmetric, but components with $|m|$\,=\,1 are spread wider (0.41505$\pm$0.00042\,$\mu$Hz on average) compared to the $|m|$\,=\,2 modes (0.41290$\pm$0.00044\,$\mu$Hz). This leads us to the conclusion that rotation in KIC\,10139564 may be slowest on the equator which would be opposite to the result obtained by \cite{baran09}.

Pulsation amplitudes of identified multiplet components can also be used to constrain the pulsation inclination \citep[supplementary materials of][]{charp11}. If the components of individual multiplets are intrinsically driven to the same amplitude, then observed amplitude differences are caused by the inclination at which they are viewed. This is particularly true for $l\geq3$ modes which have a higher degree of geometric cancellation \citep[Figure\,8.5 of supplementary material of][]{charp11}. For KIC\,10139564 we detected four possible triplets (some with components below the detection threshold) yet they do not have consistent amplitudes between the various components. This is particularly evident for the complete quintuplet where the $m\,=\,+1$ and $+2$ components have triple and double the amplitudes of their counterparts, respectively. Likewise, we observe a triplet with spacings which indicate we are seeing the $m\,=\,-2,\,+1$ and $+2$ components, which cannot be attributed to viewing inclination. As such, the amplitudes are not sufficiently consistent to be used for constraining the pulsation inclination.

\subsection{Companions to KIC\,10139564}
\cite{silv07} have detected the first planet around an sdB star. They showed that pulsations detected in sdBV stars can be used to search for periodic variations of pulsation phases, which could be caused by orbiting bodies. However, the pulsation phases have to be monitored for a minimum of 2 to 3 years and measured with low uncertainties. It means that resolved high-frequency and high-amplitude modes are preferred. With several high-amplitude modes, KIC\,10139564 is one of the best candidates where we can look for phase variations in terms of planet hunting. KIC\,10139564 has been observed for almost two years, but at the time of this analysis, we had access to only 15\,months of data. There are a few high-amplitude modes among the p-modes that we used in our phase analysis. Unfortunately, the highest amplitude periodicity is part of a triplet, that exhibits amplitude variation, and when the data are divided into subsets, a strong beating pattern in the phase analysis appears. There is a single periodicity f$_{\rm 18}$ with a relatively high amplitude. We have calculated amplitudes and phases of that periodicity in 30-day bins.
The phases are stable to within 1\,sec except for three points deviating by 5$\sigma$. The uncertainties of the phases are of the order of 0.5\,sec. A similar flat trend was found using components of the quintuplet at 5287\,$\mu$Hz but with correspondingly larger uncertainties. It is beneficial to continue monitoring these periodicities for the remainder of the {\it Kepler} mission which should determine whether a companion (planet) exists. Spectroscopically, KIC\,10139564 appears as a single star.

\subsection{Exotic species in the zoo}
\label{exotic}
Several regions in the amplitude spectrum appear to be frequency-crowded, unresolved, or ``forked'' which could indicate amplitude and/or phase variability. Investigating these regions has yielded unanticipated results, adding to the diversity of the pulsation zoo within KIC\,10139564.

\begin{figure*}
\includegraphics[width=120mm,angle=-90]{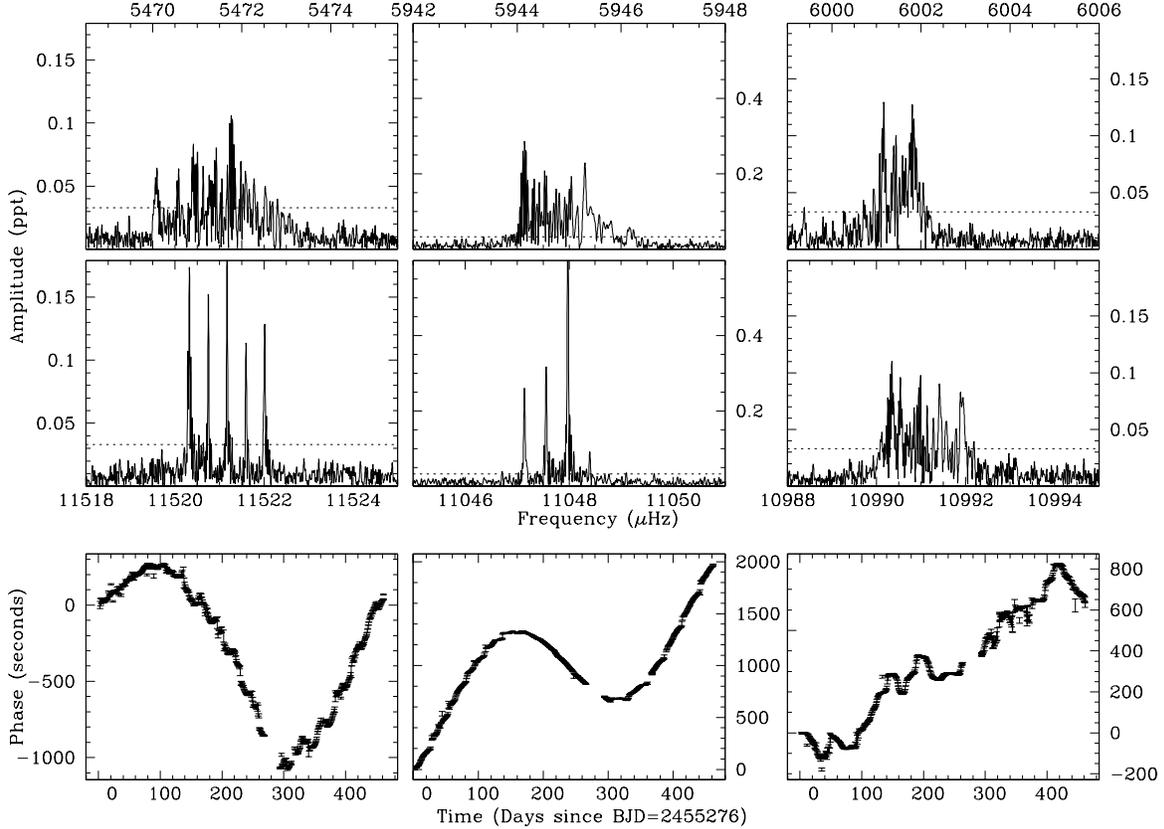}
\caption{The frequency regions near 5471, 5945 and 6003\,$\mu$Hz. The top panel shows the subNyquist regions, the middle panels the corresponding superNyquist regions and the bottom panel the phases of the subNyquist frequencies with errorbars.}
\label{figNY}
\end{figure*}

The first exotic result is that of the regions around 5472 and 5645\,$\mu$Hz. While the region looks quite messy in the combined amplitude spectrum (top panels of Figure\,\ref{figNY}), the phases (shown in the bottom panels of Figure\,\ref{figNY}), calculated using five-day sliding bins, drifted in a smooth fashion at a period of approximately one year. As such, we suspected an artifact related to the spacecraft but detected something even more interesting. The periodicities actually occur above the Nyquist frequency so are reflected across it to the regions listed above. The true frequencies are centered around 11521 and 11047\,$\mu$Hz. Careful examination reveals these frequencies are, in fact, combinations from high-amplitude frequencies in the subNyquist region. For the periodicities near 11521\,$\mu$Hz, there is a quintuplet formed from combinations of the main triplet (f$_{\rm 33}$ to f$_{\rm 35}$) near 5760\,$\mu$Hz.
Similarly, the quintuplet near 11047\,$\mu$Hz is caused by combinations of that same triplet (f$_{\rm 33}$ to f$_{\rm 35}$) with the quintuplet near 5287\,$\mu$Hz (f$_{\rm 8}$ to f$_{\rm 12}$). All combination frequencies we detected are listed in Table\,\ref{combination}. While they are reflected across the Nyquist, they become smeared by the spacecraft's motion. The exposure cadence is fixed in the spacecraft reference frame. However, the spacecraft is moving relative to the observed field, so the Nyquist frequency varies in the barycentric reference frame; this movement leads to a smearing of all frequencies whenever they are reflected across the Nyquist frequency. Therefore, these two messy regions are, in fact, reflections of combination frequencies above the Nyquist frequency.

\begin{table}
\centering
\caption{The list of combination frequencies we found in the superNyquist region. We fitted these frequencies independently from those listed in Table\,\ref{10139564list}.}
\label{combination}
\begin{tabular}{@{}ccccc@{}}
\hline \hline
ID   & Frequency [$\mu$Hz]  & Period[s]       & Amplitude [ppt] & S/N \\
\hline\hline
f$_{\rm C1}$ &11046.7169(25) &     90.524634(21) &   0.042(8) &   5.1 \\
f$_{\rm C2}$ &11047.13862(42) &     90.5211779(34) &   0.251(8) &  30.2 \\
f$_{\rm C3}$ &11047.55681(34) &     90.5177513(28) &   0.310(8) &  37.2 \\
f$_{\rm C4}$ &11048.3998(11) &     90.510845(9) &   0.093(8) &   90.6 \\
f$_{\rm C5}$ &11047.97594(14) &     90.5143173(11) &   0.754(8) &  11.2 \\
f$_{\rm C6}$ &11520.3240(6) &     86.8031145(45) &   0.175(8) &  21.1 \\
f$_{\rm C7}$ &11520.7520(7) &     86.799890(5) &   0.153(8) &   18.4 \\
f$_{\rm C8}$ &11521.1716(6) &     86.7967284(43) &   0.185(8) &  22.3 \\
f$_{\rm C9}$ &11521.5978(10) &     86.793517(7) &   0.110(8) &   13.2 \\
f$_{\rm C10}$ &11522.0152(8) &     86.790374(6) &   0.126(8) &   15.1 \\
\hline \hline
\end{tabular}
\end{table}

\begin{figure}
\includegraphics[width=83mm]{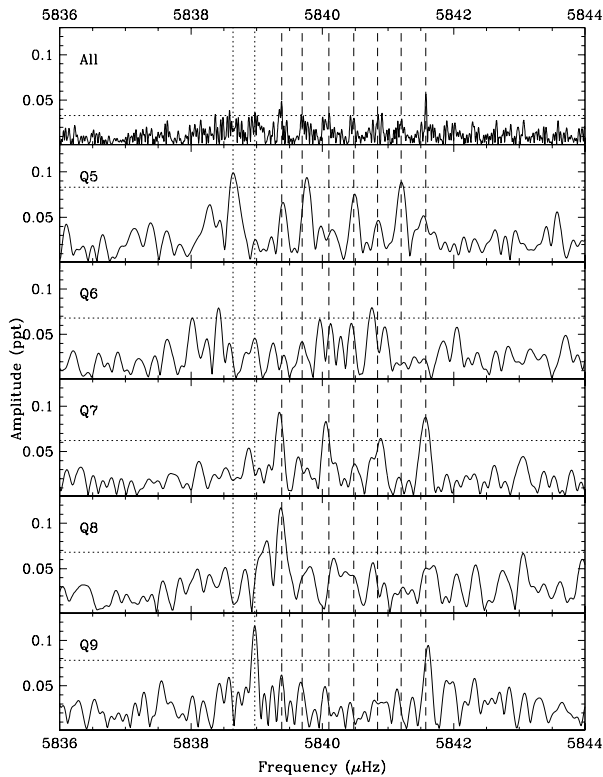}
\caption{The frequency region near 5840\,$\mu$Hz. The top panel shows the entire 15-month data set, while subsequent panels show Q5 through Q9 individually. The vertical dashed lines indicate frequencies detected using Q5 through Q8 data, and the dotted lines indicate two additional frequencies above the detection threshold in individual quarters that match the splitting. The horizontal lines denote the detection thresholds.}
\label{q5840}
\end{figure}

For other regions, this approach did not yield results, as shown in the right panels of Figure\,\ref{figNY}. Our best solutions produced another mixed set of exotic results. Amplitude variability has been previously observed in many sdBV stars \citep{reed07}, so it is not surprising also to encounter it here. One solution, therefore, is to investigate shorter sets of data over which the amplitudes and/or phases may be fairly stable. Our first approach was to divide the data into quarters, which is how the data are released to us. Figure\,\ref{q5840} shows a region where this method resulted in a solution. The region around 5840\,$\mu$Hz was known to have portions of multiplets from previous quarters of data. Using Q5 through Q8 data, the amplitude spectrum contained a complete septuplet above the detection threshold. The peaks were not symmetric, indicating phase or amplitude variation, and when Q9 was added, several peaks diminished below the detection limit. This sort of behavior is common for pulsations which are damped and re-excited. In Figure\,\ref{q5840}, the septuplet frequencies of Q5 through Q8 are marked with dashed lines. The quarterly amplitude spectra reveal that \emph{no} frequencies are above the detection limit during all five quarters, yet this region is quite frequency-rich. Quarters 5, 7 and 9 all show several substantive peaks, which fit the previously established multiplet splittings, and dotted lines in Figure\,\ref{q5840} indicate two previously unrevealed frequencies. The average spacing between all these peaks is 0.37\,$\mu$Hz. The splittings are not identical, and there is likely some frequency wandering between quarters. This may be caused by amplitude variations and/or phase interactions with other nearby periodicities which are also amplitude variable. We cannot claim all these periodicities are intrinsic to the star, but the evidence suggests this region contains an $l\geq$\,4 mode. We also examined the multiplets near 3540 and 5049\,$\mu$Hz by quarter, but no peaks were detected except those listed in Table\,\ref{10139564list}.

\begin{table}
\centering
\caption{Frequencies detected from analyses discussed in \S\ref{exotic}. This list also includes prewhitened frequencies from Table\,\ref{10139564list}.}
\label{mults}
\begin{tabular}{@{}ccccc@{}}
\hline \hline
ID & Frequency [$\mu$Hz] & $\Delta\nu$ & $l$&m \\
\hline \hline
X$_{\rm 1}$ & 5838.64 & X$_{\rm 2}$-0.33 & &-4\\
X$_{\rm 2}$ & 5838.97 & f$_{\rm 36}$-0.41 & &-3\\
f$_{\rm 36}$ & 5839.380 & X$_{\rm 3}$-0.31 & &-2\\
X$_{\rm 3}$ & 5839.694 & X$_{\rm 4}$-0.41 & &-1\\
X$_{\rm 4}$ & 5840.100 & -- & 4&0\\
X$_{\rm 5}$ & 5840.479 & X$_{\rm 4}$+0.38 & &+1\\
X$_{\rm 6}$ & 5840.841 & X$_{\rm 5}$+0.36 & &+2\\
X$_{\rm 7}$ & 5841.198 & X$_{\rm 6}$+0.36 & &+3 \\
f$_{\rm 37}$ & 5841.577 & X$_{\rm 7}$+0.38 & &+4\\
\hline
f$_{\rm 38}$ & 6001.10 & f$_{\rm 39}$-0.37 & &-1\\
f$_{\rm 39}$ & 6001.47 & -- & 1&0\\
X$_{\rm 8}$ & 6001.88 & f$_{\rm 39}$+0.41 & &+1\\
\hline
X$_{\rm 9}$ & 6057.4 & -- & --\\
X$_{\rm 10}$ & 6057.6 & -- & -- \\ \hline
X$_{\rm 11}$ & 6170.83 & X$_{\rm 12}$-0.34 & &-1\\
X$_{\rm 12}$ & 6171.17 & --  & 1&0\\
X$_{\rm 13}$ & 6171.66 & X$_{\rm 12}$+0.49 & &+1\\
\hline
X$_{\rm 14}$ & 6172.39 & X$_{\rm 15}$-0.44 & &-1\\
X$_{\rm 15}$ & 6172.83 & -- & 1&0\\
X$_{\rm 16}$ & 6173.32 & X$_{\rm 15}$+049 & &+1\\
\hline
X$_{\rm 17}$ & 6234.3 & f$_{\rm 47}$-0.39 & &-1\\
f$_{\rm 47}$ & 6234.7 & -- & 1&0\\
f$_{\rm 48}$ & 6235.05 & f$_{\rm 47}$+0.32 & &+1\\
\hline \hline
\end{tabular}
\end{table}

\begin{figure*}
\includegraphics[width=90mm,angle=270]{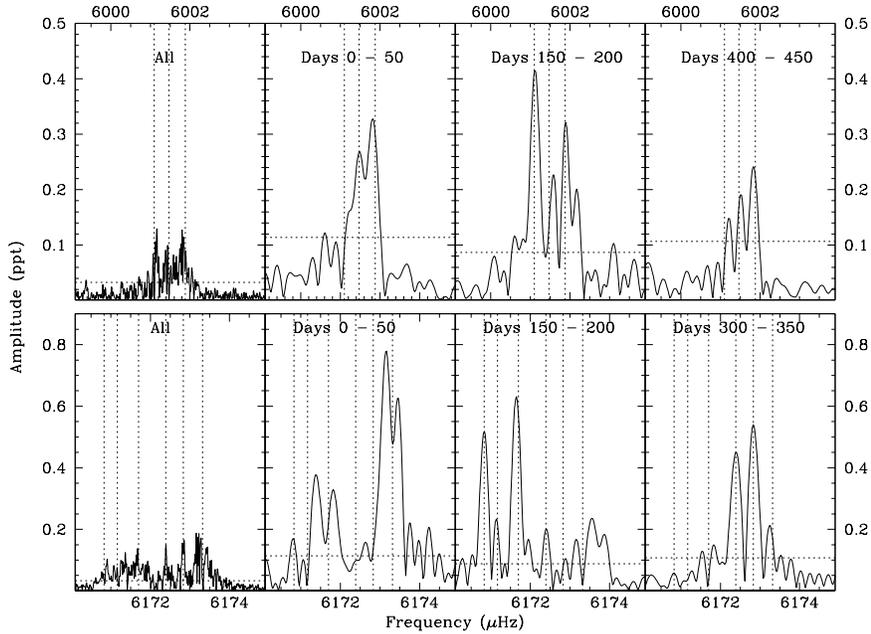}
\caption{Amplitude spectra for the regions around 6001 and 6173\,$\mu$Hz. Left panels show the entire data set, while the remaining panels show 50-day segments as labelled. Dotted lines indicate likely frequencies listed in Table\,\ref{mults}.}
\label{figmessy}
\end{figure*}

If quarterly data did not provide a solution for a given region, we examined them using shorter data sets. We calculated amplitude spectra for sets of data spanning 50, 60 and 70 days, avoiding the gap between days 270 and 290. The top panels of Figure\,\ref{figmessy} and Table\,\ref{mults} show our solution for the region near 6001\,$\mu$Hz. There appears to be some frequency wandering, but from these subsets of data, we estimate frequencies lie near 6001.10, 6001.47 and 6001.88\,$\mu$Hz, which give an overall separation of 0.39\,$\mu$Hz, roughly the established rotational multiplet splitting. Most likely, this region contains a triplet with a separation of the spin period.

The situation near 6173\,$\mu$Hz is more complex (Figure\,\ref{figmessy}), largely because the amplitudes become very low between days 130 to 220 and after day 330. Additionally, there are more frequencies involved in this region which covers $\sim3\mu$Hz. It appears there is a triplet near 6170.83, 6171.17 and 6171.66\,$\mu$Hz, which switches power from the right pair to the center frequency to the outside pair prior to day 150. Then another triplet, which is modest at the beginning, gets even weaker in the middle but comes back at the end of the run. These frequencies are near 6172.39, 6172.83 and 6173.32\,$\mu$Hz, which would be another triplet. However, these cannot both be $l$\,=\,1 triplets, since they are too close together, so one of them is likely a part of an $l$\,=\,2 multiplet. This region, which is shown in the bottom panels of Figure\,\ref{figmessy}, is very complex, and we can only roughly estimate the frequencies of Table\,\ref{mults}. Any interpretations are fairly speculative. We estimate that this region contains two distinct triplets with splittings of the rotation period, but we certainly cannot rule out a single multiplet.

\begin{figure*}
\includegraphics[width=90mm,angle=270]{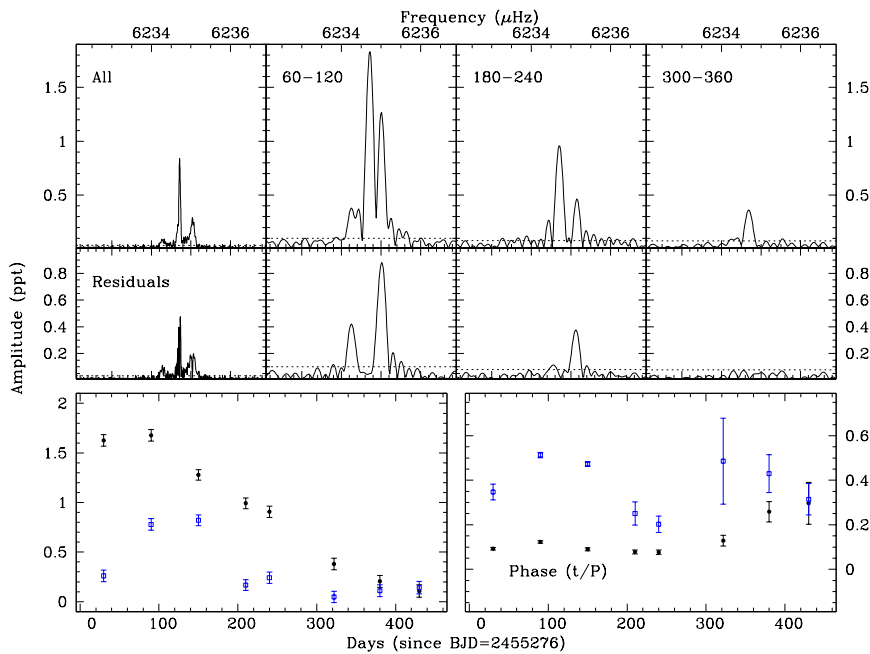}
\caption{The frequency region near 6235\,$\mu$Hz. Original amplitude spectra (top), residuals after prewhitening the strongest frequency (middle) and the amplitudes (bottom left) and phases (bottom right) for the frequencies at 6234.7 (filled circles) and 6235.0\,$\mu$Hz (open squares).}.
\label{fig6235}
\end{figure*}

The last region we discuss is near 6235\,$\mu$Hz (Figure\,\ref{fig6235}). In this region, there is a high amplitude peak in the overall amplitude spectrum with two sidelobes, none of which prewhiten well. Breaking the data into 60-day subsets to resolve the frequencies clarifies what is occurring. In this case the main frequency f$_{\rm 42}$ at 6234.7 is frequency and phase stable with an amplitude that diminishes after day $\sim$80. There are two other periodicities, both with frequencies and amplitudes that vary during the observations. The higher frequency (f$_{\rm 43}$) drifts from 6235.07 to 6235.02\,$\mu$Hz and back again during the run, while the lower frequency drifts from 6234.4 to 6234.2\,$\mu$Hz prior to day 180, after which it is no longer detectable. This is shown in Figure\,\ref{fig6235} where the original amplitude spectra are shown in the top panels, the amplitude spectra prewhitened by f$_{\rm 42}$ in the middle panels and the amplitudes and phases in the bottom panels. Clearly the amplitudes of all frequencies are variable. The amplitudes for f$_{\rm 48}$ are only significant for the fits centered at days 90 and 150 while the amplitude for f$_{\rm 47}$ continually declines until by day 380 it is no longer significant.

We also attempted to disentangle the region near 6314\,$\mu$Hz using these techniques, but we were not able to establish reasonably stable frequencies. The estimated frequencies for these regions appear in Table\,\ref{mults}.

\section{Summary}
KIC\,10139564 is the only p-mode dominated sdBV star observed by {\it Kepler}. It was discovered during the survey phase of the mission and has been observed continuously in SC mode since the start of the second year. The SC mode allows us to study short period pulsations that would be unobservable with LC sampling. KIC\, 101539564 contains frequencies higher than the Nyquist frequency associated with the SC sampling interval, but thanks to the long duration of these observations the barycentric correction induces frequency modulations that resolves the ambiguity of modes reflected across the Nyquist frequency into the subNyquist region.

The amplitude spectrum  calculated from 15\,months of nearly continuous data shows more than 50 periodicities in a pulsation zoo, which includes both anticipated and exotic species. We detected the anticipated p-modes, found several g-modes and two periodicities between the p- and g-mode regions. These possibly mixed modes have been observed in some other sdBV stars \citep{reed10,baran11a}, and are particularly abundant in the complex pulsator 2M1938+4603 \citep{osten10c}. The highest amplitude is around 8\,ppt, and only a few other frequencies are of comparable height. Most of the signal is below 0.5\,ppt, which would be difficult to detect from the ground.

Almost all periodicities are split into frequency multiplets. We interpret this splitting as caused by rotation and derive an average rotation period of 25.6$\pm$1.8\,days. The observed splittings vary by a small amount, leading to an uncertainty in the rotation period. This is most likely caused by small differences in the Ledoux constant as has been observed in models calculated by \citet{charp02}. The scale of variation in the Ledoux constant matches the variation in frequency spacings. The exact rotation rate can be established once the order {\it n} and degree $l$ are identified to better constrain the Ledoux constant. Most gravity modes appear as triplets, and their frequency differences are consistent with identifying them as $l$\,=\,1 modes.

They also show nearly equal period spacings, which we interpret as asymptotic overtones of a single degree. Multiplets indicate they are $l$\,=\,1 modes, but the period spacings are too large to be of consecutive overtones. They are more likely 207 or 310\,s which is dissimilar to the 250\,s observed for the other \emph{Kepler}-observed pulsating sdB stars \citep{reed11}. However, those other stars are all g-mode dominated and so are cooler than KIC\,10139564. The only other sdB star that is a p-mode dominated hybrid is Bal09. Ground-based observations detected 20\,g-mode periodicities yet there is no indication of consistent period spacings between them. So at this time Bal09 is the only pulsating sdB star with rich g-mode spectrum which does not show asymptotic g-mode period spacings. It is also both cooler and more evolved (lower $\log g$) than KIC\,10139564 and so may not be any more suitable for comparison than the other g-mode pulsators. Theoretical works in progress (by VVG) are seeking to explain this discrepancy.


Several of the multiplets show signal excesses at the frequencies of likely missed components and are indicated in the figures with vertical dash-dotted lines. We included them in our list as the S/N\,=\,4 threshold is adopted to separate intrinsic peaks from random noise. When there is good reason to anticipate pulsation frequencies, in this case multiplet splitting, the significance threshold does not necessarily apply. We do not claim that any of these frequencies are intrinsic to the star, yet we include this speculation to compel future investigation with additional data.

Recent spectroscopic measurements indicate KIC\,10139564 is not a radial velocity variable and allowed us to refine its temperature and gravity, placing it well away from the typical hybrid pulsator region \citep[see Fig.\,2 of][]{osten10b}. This raises the possibility that low-amplitude, g-mode pulsations are common place in V361\,Hya class stars, but an instrument with {\it Kepler's} accuracy is required to observe them. A re-examination of pulsation driving models would seem necessary as the most recent g-mode driving publications \citep{jeffery06,hu11} cannot drive pulsations to the temperature of KIC\,10139564. Improvements for diffusion of iron-group elements (in progress by HH) seems likely to solve this issue.


Among the p-modes at high frequencies, we have noticed several regions to be unresolved in frequencies, and our first guess led us towards stochastic oscillations as a possible explanation. Detailed analysis showed, however, that they are likely not stochastic in nature, but rather require multiple explanations. The regions all appear as multiplets and frequency splittings are consistent with our derived rotation period. Two regions are caused by reflections across the Nyquist frequency of combination frequencies, which are smeared by the difference in spacecraft and barycentric reference frames.  Others are caused by highly variable amplitudes, and some appear to be changing their frequencies slightly. In the combined data set, this leads to multiple-peaked or even transient periodicities with amplitudes that are greatly reduced.

As a result of the abundant multiplets, we were able to associate most frequencies with modes, which is a very rare result for sdBV stars. Also of interest are the probable detections of $l$\,=\,3 or greater modes. In many other sdBV stars, frequency density has strongly suggested that higher $l$ modes are excited \citep{reed07}, yet because of geometric cancellation, there has been some debate whether we would most likely observe $l$\,=\,3 or 4 modes. Mode density in this star also creates a new problem because current models \citep[\emph{e.g.}][]{charp02} cannot produce as many modes as are observed. From a sample of representative models, the overtone spacings for $l$\,=\,0 to 4 averaged from 904 to 939\,$\mu$Hz. Near the fundamental order, this could be reduced to about 700\,$\mu$Hz.
For the better-established sequences, we detect overtone separations of 237, 124, 176, 241 and 234$\mu$Hz. Such a dense spectrum is clearly challenging for current models yet could go a long way to explain the high-density observed in several sdBV stars. KIC\,10139564 seems well placed to solve these issues, and the many multiplets (especially those of higher degree) should provide constraints on the rotation axis as well.

Only one high-amplitude p-mode is detected to be a single peak. Its amplitude is almost 4\,ppt. We used this peak to match amplitudes between months to reduce residual signal as much as possible. This helped to diminish residuals for most of the peaks, which could mean either the change of intrinsic amplitudes of those periodicities is consistent or they are stable and any difference in amplitude between quarters is caused by systematics.

Fifteen months of continuous data coverage opened a way to study phase stability of detected periodicities to search for possible companions to KIC\,101039564. Although we have detected more than 50 periodicities, we could not use most of them for checking phase stability. Close neighbouring peaks or residuals, which indicate amplitude/phase variation, cause a beating pattern in both amplitude and phase. We found only one well-suited frequency around 5472\,$\mu$Hz, which we used to look for phase stability. It is still affected by some nearby low-amplitude, complex signals. However, the beating between the peak and that complex signal is about 10\,days, so the beat period is shorter than the period from rotational splitting. As a result of our phase analysis, we have not noticed any periodic variation in phase that could be interpreted as a companion to KIC\,10139564. There seems to be a real deviation in phase between day 420 and 520, but we are not yet able to interpret this variation.

The results we presented in this paper clearly show KIC\,10139564 to be very interesting and allowed us to detect features not accessible from ground-based data. As most frequency regions contained multiplets, mode identifications could be determined for most of the peaks we detected. The identified modes should pose strong constraints on stellar models and should, obviously, allow the exclusion of non-unique solutions, making the theoretical study of this object easier. In fact, this is the only sdBV star with so many modes identified. As a natural implication of the pulsation variety and all the remarkable results we obtained from only 15\,months of data, we definitely should continue monitoring this star to confirm several features as well as to uncover more properties still hidden in the data.

\section*{Acknowledgments}

AB gratefully appreciates funding from the Polish Ministry of Science and Higher Education under project No. 554/MOB/2009/0. MDR was supported by the Missouri Space Grant Consortium, funded by NASA. The research leading to these results has received funding from the European Research Council under the European Community's Seventh Framework Programme (FP7/2007--2013)/ERC grant agreement N$^{\underline{\mathrm o}}$\,227224 ({\sc prosperity}), as well as from the Research Council of K.U.Leuven grant agreement GOA/2008/04. PD, SB, HH and SK are grateful for the hospitality during their stay at the Kavli Institute for Theoretical Physics in the framework of the Research Program Asteroseismology in the Space Age. We thank U.\,Heber for kindly providing the spectral model grids. The authors thank Tracie Dalton for improving the manuscript by making English correction. Funding for the {\it Kepler} Mission is provided by NASA's Science Mission Directorate. The authors gratefully acknowledge the {\it Kepler} Science Team and all those who have contributed to making the {\it Kepler} Mission possible.

\section*{Author Contributions}
ASB prepared the manuscript, analyzed and interpreted the data from which the list of periodicities was inferred, identified multiplets and discussed the results. MDR contributed through \S\ref{exotic} and discussed the results. RH{\O} and JHT obtained and analyzed spectroscopy, supplied the manuscript with \S\ref{spectroscopy} and discussed the results. DS, EP, SJO and PD checked the stability of periodicities in \S\ref{exotic} and discussed the results. RS contributed to discussions on mode stability. SB and HH supplied us with evolutionary models. VVG commented on evolutionary models. BDC, JVC, SET contributed to the {\it Kepler} mission. SK contributed to the overall discussion.

\label{lastpage}

\end{document}